\documentclass[aps,prl,twocolumn,showpacs,superscriptaddress,groupedaddress,nofootinbib]{revtex4}  
\usepackage{graphicx}  
\usepackage{dcolumn}   
\usepackage{bm}        
\usepackage{amssymb}   
\usepackage{amsmath}
\usepackage{float}
\usepackage{slashed}
\usepackage{hyperref}
\usepackage{color} 
\usepackage{bbold}
\usepackage{multirow}

\newcommand{\code}[1]{\texttt{#1}}

\def\B0{B^{(0)}}
\def\A0{A_3^{(0)}}

\def\amu{$\Delta{a^{\rm FB}_\mu}$}

\begin{document}

\hspace{2.1in}
MS-TP-21-10
\vspace{-0.5cm}

\title{What Fermilab $(g-2)_{\mu}$ experiment  tells us about discovering  SUSY at high luminosity LHC and high energy LHC}

\author{Amin Aboubrahim}
\thanks{{\scriptsize Email}: \href{mailto:aabouibr@uni-muenster.de}{aabouibr@uni-muenster.de}; {\scriptsize ORCID}: \href{https://orcid.org/0000-0002-1110-4265}{0000-0002-1110-4265}}
\affiliation{Institut f\"ur Theoretische Physik, Westf\"alische Wilhelms-Universit\"at M\"unster, Wilhelm-Klemm-Stra{\ss}e 9, 48149 M\"unster, Germany}
\author{Michael Klasen}
\thanks{{\scriptsize Email}: \href{mailto:michael.klasen@uni-muenster.de}{michael.klasen@uni-muenster.de}; {\scriptsize ORCID}: \href{https://orcid.org/0000-0002-4665-3088}{0000-0002-4665-3088}}
\affiliation{Institut f\"ur Theoretische Physik, Westf\"alische Wilhelms-Universit\"at M\"unster, Wilhelm-Klemm-Stra{\ss}e 9, 48149 M\"unster, Germany}
\author{Pran Nath}
\thanks{{\scriptsize Email}: \href{mailto:p.nath@northeastern.edu}{p.nath@northeastern.edu}; {\scriptsize ORCID}: \href{https://orcid.org/0000-0001-9879-9751}{0000-0001-9879-9751}}
\affiliation{Department of Physics, Northeastern University, Boston, MA 02115-5000, USA}
\date{\today}
\begin{abstract} 
Using an artificial neural network we explore the parameter space  of supergravity grand unified models 
 consistent with the combined  Fermilab E989  and Brookhaven E821 
  data on $(g-2)_\mu$. Within an extended mSUGRA model with non-universal gaugino masses
 the analysis indicates that the region favored by the data  is the one generated by gluino-driven radiative breaking of the electroweak symmetry ($\tilde g$SUGRA). This region naturally leads to a split sparticle spectrum with light sleptons and weakinos but heavy squarks, with the stau and the chargino as the lightest charged particles. We show that if the entire deviation from the Standard Model  $(g-2)_{\mu}$   arises from supersymmetry, then 
  supersymmetry is discoverable at  HL-LHC and HE-LHC
  via production and decay of sleptons and sneutrinos within the optimal integrated luminosity of HL-LHC and with 
  a smaller integrated luminosity at HE-LHC. The effect of CP phases on the muon anomaly is 
  investigated and the parameter space of CP phases excluded by the Fermilab constraint is exhibited.
  
 \end{abstract}

\maketitle

\section{1. Introduction \label{1}}

Recently the Fermilab E989 experiment has measured $a_\mu=(g-2)_\mu/2$ with an unprecedented accuracy 
so that~\cite{Abi:2021gix} 
 \begin{equation}
 a^{\rm exp}_\mu =   116 592 040 (54)  \times10^{-11} 
 \quad \quad \text{(Fermilab E989)}.
  \label{fermi}
\end{equation} 
This is to be compared with the previous Brookhaven experiment E821~\cite{Bennett:2006fi,Tanabishi}
 which gave 
 \begin{equation}
 a^{\rm exp}_\mu = 116 592 091 (63)  \times10^{-11} \quad~~\text{(Brookhaven E821)}. 
  \label{Brook}
\end{equation} 
The combined Fermilab and Brookhaven   data give 
 \begin{equation}
 a^{\rm exp}_\mu = 116 592 061 (41)  \times10^{-11} \quad\text{ (Combined  E989+E821)}.
  \label{Brook}
\end{equation} 
The combined result is to be compared with the Standard Model (SM) prediction which gives~\cite{Aoyama:2020ynm} 
 \begin{equation}
\ a^{\rm SM}_\mu = 116 591 810 (43)  \times10^{-11},
  \label{SM}
\end{equation} 
where the Standard Model prediction contains precise quantum electrodynamic, electroweak, hadronic 
vacuum polarization and hadronic light-by-light contributions. Thus the difference between the combined
Fermilab and Brookhaven (FB) result and the SM result is 
  \begin{equation}
\Delta a^{\rm FB}_\mu = a_{\mu}^{\rm exp} - a_{\mu}^{\rm SM} = 251 (59)  \times10^{-11},
  \label{diff}
\end{equation} 
which is a $4.2\sigma$ deviation of experiment from the SM result. 
Eq.~(\ref{diff}) confirms the Brookhaven result of a discrepancy and further strengthens it, i.e., 4.2$\sigma$ 
vs $3.7\sigma$ for Brookhaven.  Although not yet a discovery of new physics which requires $5\sigma$,
Eq.~(\ref{diff})  is now  more compelling than the Brookhaven result alone as a harbinger of new physics (see, however, Ref.~\cite{Borsanyi:2020mff}).

In this work we investigate if \amu~given by Eq.~(\ref{diff}) can arise from the electroweak sector 
of supersymmetric models. In the SM the electroweak corrections arise from the exchange of the
$W$ and $Z$ bosons~\cite{fuji,Czarnecki:1995sz}.
It is known from early days that supergravity (SUGRA) unified models can generate supersymmetric loop 
corrections to the muon anomaly
 from the exchange of charginos and muon-sneutrino, and from the exchange of neutralinos and smuons
 which  can be comparable to the SM electroweak corrections~\cite{Kosower:1983yw,Yuan:1984ww}.
However, the supersymmetric contribution depends sensitively on the SUGRA parameter space, 
specifically on the soft parameters~\cite{Lopez:1993vi,Chattopadhyay:1995ae,Moroi:1995yh,Carena:1996qa}
and an exploration of the parameter space is needed to satisfy the experimental constraint. Thus 
the Brookhaven experiment~\cite{Bennett:2006fi}  led to a number of
works~\cite{Czarnecki:2001pv,Chattopadhyay:2001vx,Everett:2001tq,Feng:2001tr,Baltz:2001ts}
exploring the parameter space of supersymmetry (SUSY) and supergravity models.
Since then the discovery of the Higgs boson at 125 GeV~\cite{Chatrchyan:2012ufa,Aad:2012tfa} has put further constraint on the parameter
space of SUSY models. This is so since at the tree level, supersymmetric models
imply that the Higgs boson mass lies below $M_Z$, and thus one needs a large loop correction to lift the
Higgs mass to the experimentally observed value. This in turn implies that the size of weak scale
SUSY must be large lying in the several TeV region~\cite{Akula:2011aa,higgs7tev1} which further 
restricts the supergravity models. 
In view of the experimental data from Fermilab~\cite{Abi:2021gix}, we investigate in this work the implications of the  
combined Fermilab and Brookhaven result  \amu~for supergravity models and  for discovering supersymmetry at 
  HL-LHC and HE-LHC. To this end, we carry out a comprehensive analysis of the parameter space of
 supergravity grand unified models~\cite{sugrauni} 
 using an artificial neural network (ANN) with constraints on the Higgs mass, the dark matter relic density and the muon $g-2$. Machine learning methods are found efficient when exploring 
large parameter spaces (see, e.g., Refs.~\cite{Hollingsworth:2021sii,Balazs:2021uhg}). It is observed that the allowed regions of the parameter space are those where gluino-driven radiative 
 breaking of the
 electroweak symmetry occurs~\cite{Akula:2013ioa,Aboubrahim:2019vjl,Aboubrahim:2020dqw}
 referred to as $\tilde g$SUGRA. In this region, the sleptons (selectrons and smuons), sneutrinos and the 
electroweakinos can be light while squarks and the extra Higgs bosons of the MSSM, i.e., $A^0, H^0, H^{\pm}$ are
all heavy. The  lightest charged particles are the stau, the smuon, the selectron and the chargino. Using a deep neural network (DNN), we investigate
the prospects of the discovery of sleptons  and sneutrinos at HL-LHC and HE-LHC 
in the framework of SUGRA grand
unified models assuming non-universality of gaugino masses~\cite{Ellis:1985jn,nonuni2,Feldman:2009zc,Belyaev:2018vkl}. 

The outline of the rest of the paper is as follows: In section 2 we carry out a scan of the extended mSUGRA parameter space with two additional parameters in the gaugino mass
sector. In section 3 an analysis of sparticle spectrum and dark matter constrained by \amu~is given.
In section 4 we investigate the 
implications of \amu~for discovering SUSY at HL-LHC and HE-LHC and give the estimated integrated luminosities for the benchmarks in section 5. 
In section 6 an analysis of the constraints on CP phases from \amu~is given and it is 
shown that a significant part of the parameter space is eliminated by the CP phases.
Conclusions are given in section 7.

\section{2. Scan of the constrained SUGRA parameter space \label{sec2} }

\begin{figure}[t]
\centering
\includegraphics[width=0.49\textwidth]{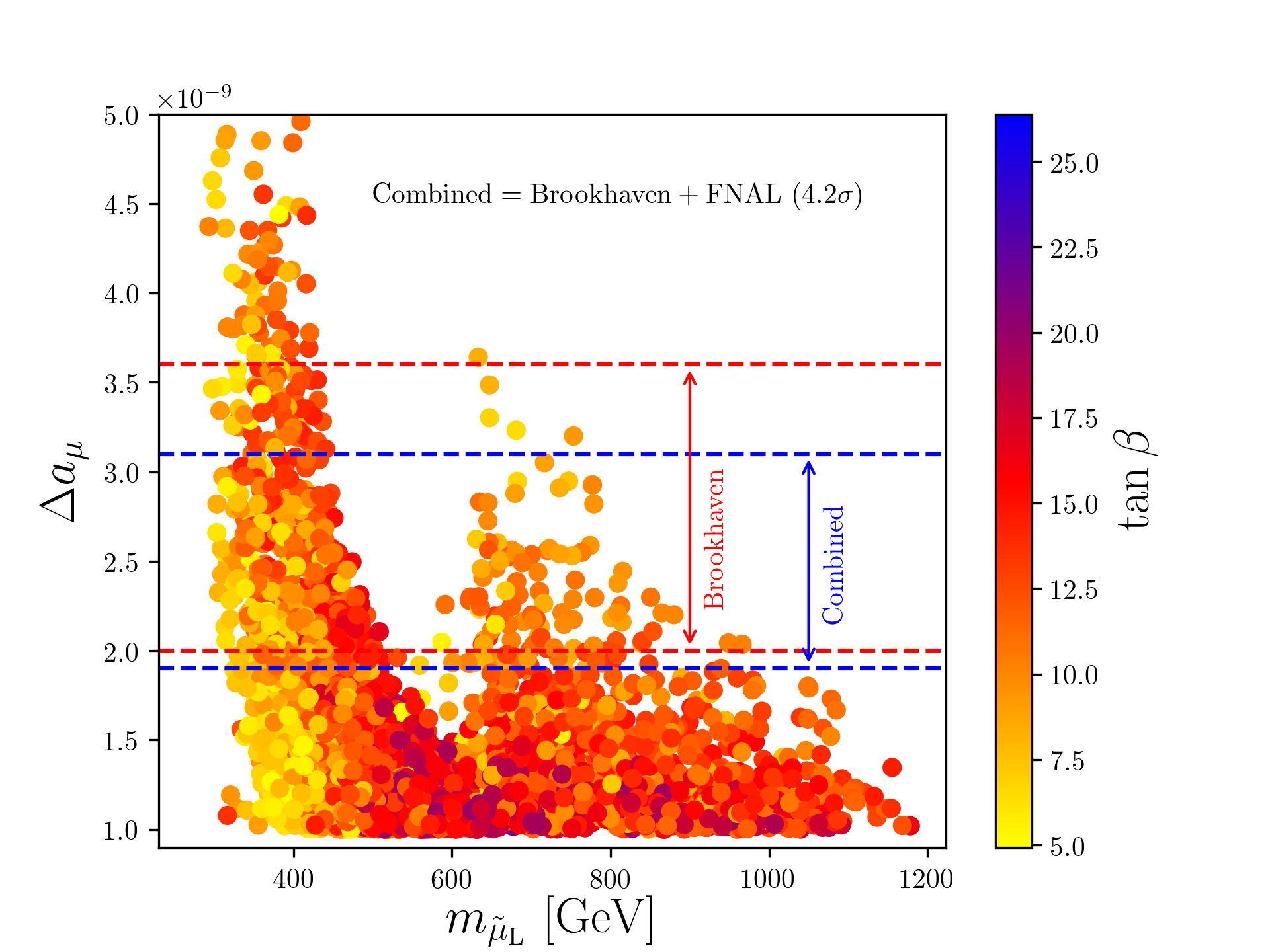}
\includegraphics[width=0.49\textwidth]{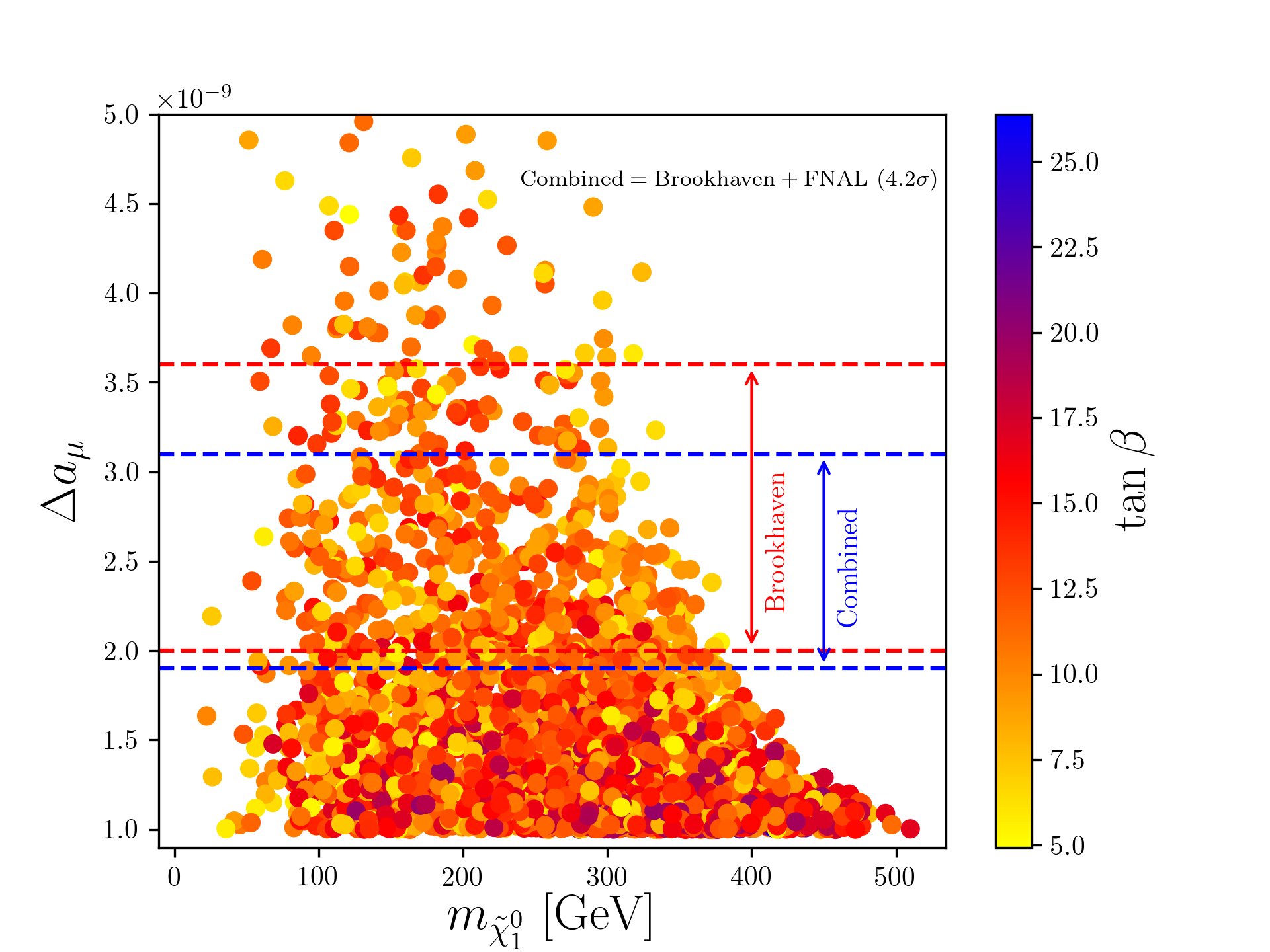}
\caption{$\Delta a^{\rm SUSY}_{\mu}$ arising from the SUGRA parameter space using  an ANN. The points satisfy the Higgs boson mass, the dark matter relic density and limits from dark matter direct detection experiments and the LHC. One sigma error corridor on $\Delta a_{\mu}$ experimental result from Brookhaven
(red dashed lines) and from the combined Fermilab and
Brookhaven result (blue dashed lines) are also displayed.}
\label{fig1}
\end{figure}

\begin{figure}[!htp]
\includegraphics[width=0.55\textwidth]{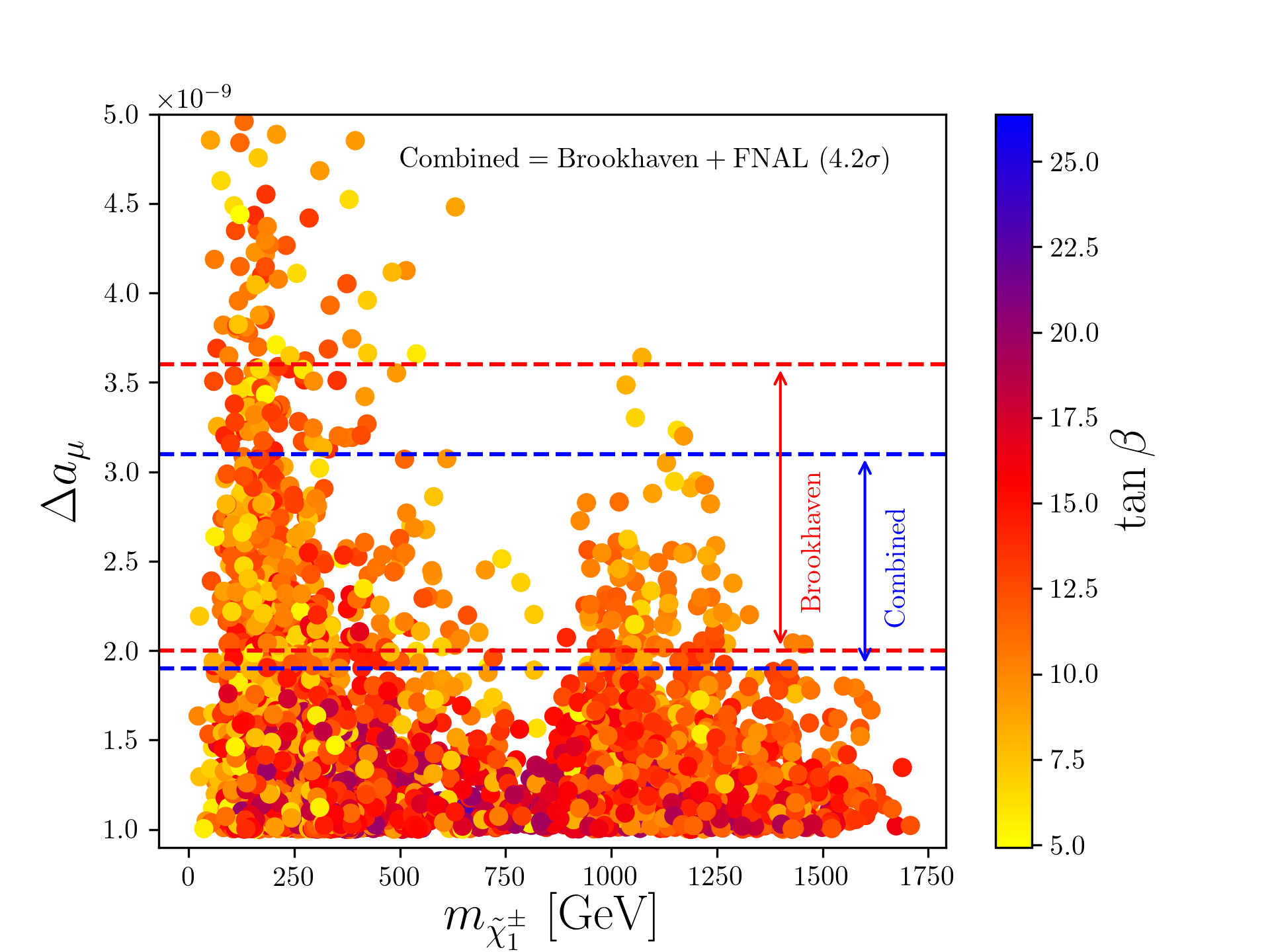}
\caption{A scatter plot of $\Delta a_{\mu}$ versus the chargino mass as a result of a scan of the SUGRA parameter space using an artificial neural network. The points satisfy the Higgs boson mass, the dark matter relic density and limits from dark matter direct detection experiments and the LHC.}
\label{fig1a}
\end{figure}

As noted earlier, the scan of the SUGRA parameter space is carried out using an artificial neural network as means to optimize the search in accordance with the most recent constraints from experiments.
Our aim is  to explore regions of the parameter space of  supergravity grand unified models  that produce a supersymmetric  loop correction $\Delta a^{\rm SUSY}_{\mu}$ 
  consistent with \amu. Thus the parameter
space of the model consists of $m_0, m_1,m_2,m_3, A_0, \tan\beta$ and sign$(\mu)$,
where $m_0$ is the universal scalar mass, 
 $m_i$ ($i$=1$-$3) are the non-universal gaugino  masses
 which are the $U(1)$, $SU(2)$ and $SU(3)$ 
gaugino masses, $A_0$ is the universal scalar coupling and $\tan\beta=v_2/v_1$, where $v_2$ 
gives mass to the up quarks and $v_1$ gives mass to the down quarks and the leptons. 
In the analysis we include the effect of two loop corrections to the $a_{\mu}^{\rm SUSY}$~\cite{Heinemeyer:2003dq}
although such corrections are typically  small. 
The scan of the parameter space uses an ANN implemented in \code{xBit}~\cite{Staub:2019xhl} interfaced with \code{SPheno-4.0.4}~\cite{Porod:2003um,Porod:2011nf} which uses two-loop MSSM RGEs and three-loop Standard Model RGEs and takes into account SUSY threshold effects at the one-loop level to generate the sparticle spectrum and \code{micrOMEGAs-5.2.7}~\cite{Belanger:2014vza} to calculate the DM relic density and the spin-independent scattering cross section. The ANN used has three layers with 25 neurons per layer. With the above constraints imposed while allowing for a $2\sigma$ window, the ANN constructs the likelihood function of a point from the three constraints and the training is done on the likelihood rather than on the observable itself. 
 The obtained set of points are then passed to \code{Lilith}~\cite{Bernon:2015hsa,Kraml:2019sis}, \code{HiggsSignals}~\cite{Bechtle:2013xfa} and \code{HiggsBounds}~\cite{Bechtle:2020pkv} to check the Higgs sector constraints as well as \code{SModelS}~\cite{Khosa:2020zar,Kraml:2013mwa,Kraml:2014sna} to check the LHC constraints. Furthermore, \code{micrOMEGAs-5.2.7}~\cite{Barducci:2016pcb} has a module which we use to check the constraints from DM direct detection experiments. The points passing all those constraints are plotted in Figs.~\ref{fig1} and~\ref{fig1a}. 

 In Fig.~\ref{fig1} we display  $\Delta a_\mu$ arising from supersymmetric loops vs the smuon mass (top panel)
 and vs the neutralino mass (bottom panel), while in  Fig.~\ref{fig1a} we show  $\Delta a_\mu$ vs the chargino mass. The $\Delta a_\mu$ constraint with a one sigma corridor arising
 from the Brookhaven experiment and from the combined Fermilab and Brookhaven data are indicated
 and it is seen that the SUGRA model points populate the region allowed by the combined data constraint.
 In the analysis here we have not taken into account SUSY CP phases but we note  in passing 
that SUSY CP  phases can have significant effect
on the supersymmetric loops corrections~\cite{Aboubrahim:2016xuz}. The constraints on SUSY phases arising from 
 \amu~are discussed later.

\begin{table}[t]
\caption{\label{tab1}
Input parameters for the benchmarks used in this analysis. All masses are in GeV.} 
\begin{ruledtabular}
\begin{tabular}{ccccccc}
Model & $m_0$ & $A_0$ & $m_1$ & $m_2$ & $m_3$ & $\tan\beta$ \\
   \hline\rule{0pt}{3ex}
\!\!(a) & 460 & -1209 & 726 & 378 & 5590 & 7.0 \\
   (b) & 685 & 1380 & 868 & 493 & 8716 & 13.0 \\
 (c) & 682 & 3033 & 875 & 714 & 8929 & 13.0 \\
 (d) & 389 & 122 & 649 & 377 & 4553 & 8.2 \\
 (e) & 254 & 1039 & 793 & 1477 & 8508 & 10.8
\end{tabular}
\end{ruledtabular}
\end{table}

\begin{table}[t]
\caption{\label{tab2}
The light sparticle spectrum consisting of the left and right handed sleptons ($\tilde\ell_{\rm L,R}$), the sneutrino ($\tilde \nu_{\rm L}$), 
 neutralino ($\tilde \chi_1^0$), chargino ($\tilde \chi_1^{\pm}$) contributing to the muon $g-2$, the DM relic density and $\Delta a_{\mu}(\times 10^{-9})$ (calculated at the two-loop level using~\code{GM2Calc}~\cite{Athron:2015rva}). Also given are the SM-like Higgs mass, the spin-independent $(\times 10^{-51})$ and spin-dependent $(\times 10^{-49})$ proton-neutralino cross-section in units of cm$^2$.}
\begin{ruledtabular}
\begin{tabular}{ccccccccccc}
Model & $h^0$ & $\tilde\ell_{\rm L}$ & $\tilde\ell_{\rm R}$ & $\tilde\nu_{\rm L}$ & $\tilde\chi^0_1$ & $\tilde\chi^{\pm}_1$ & $\Delta a_{\mu}$ & $\Omega h^2$ & $\sigma^{\rm SI}$ & $\sigma^{\rm SD}$ \\
\hline\rule{0pt}{3ex}
\!\!(a) & 123.7 & 313 & 542 & 304 & 222.2 & 222.4 & 2.13 & 0.001& 5.17 & 9.13 \\
(b) & 124.6 & 412 & 761 & 405 & 271.7 & 271.9 & 3.04 & 0.002 & 1.62 & 5.27 \\
(c) & 124.6 & 501 & 758 & 495 & 331.0 & 465.0 & 2.02 & 0.055 & 4.73 & 9.63 \\
(d) & 123.4 & 305 & 463 & 295 & 237.4 & 237.6 & 2.33 & 0.002 & 13.0 & 49.2 \\
(e) & 123.7 & 721 & 422 & 716 & 300 & 1143 & 2.56 & 0.052 & 6.34 & 9.42 
\end{tabular}
\end{ruledtabular}
\end{table}

\begin{table}[t]
\caption{\label{tab2a}
The branching ratios for the dominant decay channels of the left and right handed sleptons along with the sneutrino.}
\begin{ruledtabular}
\begin{tabular}{ccccc}
Model & $\tilde\ell_{\rm L}\to\ell\tilde\chi^0_1$ & $\tilde\ell_{\rm R}\to\ell\tilde\chi^0_1~[\tilde\chi^0_2]$ & $\tilde\nu_{\rm L}\to\tilde\chi^+_1\ell^-$ & $\tilde\nu_{\rm L}\to\tilde\chi^0_1\nu_{\ell}$ \\
\hline\rule{0pt}{3ex}
\!\!(a) & 33\% & - [100\%] & 66\% & 33\% \\
(b)  & 32\% & - [100\%] & 64\% & 32\% \\
(c) & 65\% & 100\% [-] & 20\% & 71\%\\
(d) & 31\% & - [100\%] & 62\% & 30\%\\
(e) & 100\% & 100\% [-] & - & 100\%
\end{tabular}
\end{ruledtabular}
\end{table}

\begin{figure}[!htp]
\includegraphics[width=0.55\textwidth]{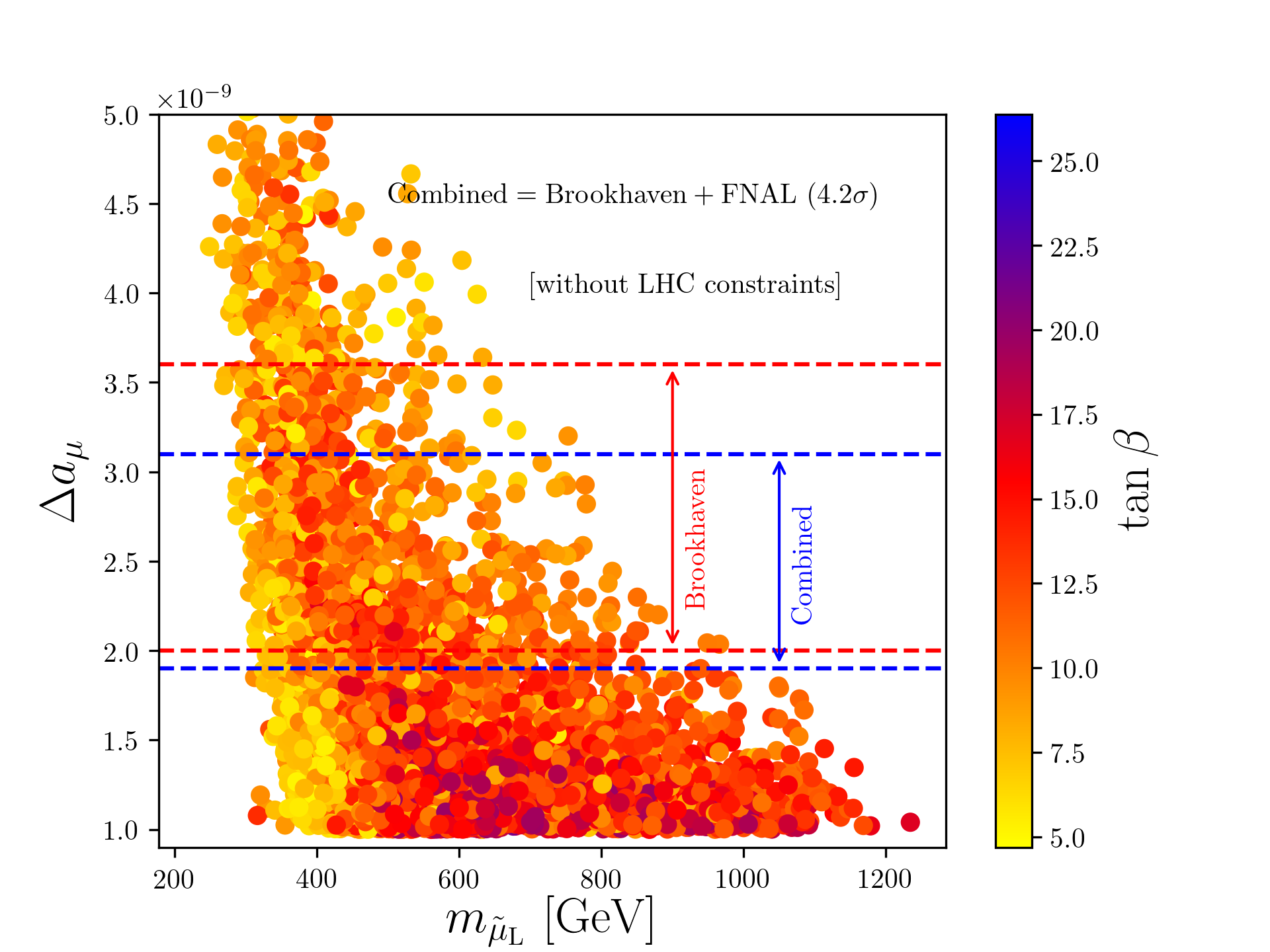}\\
\includegraphics[width=0.55\textwidth]{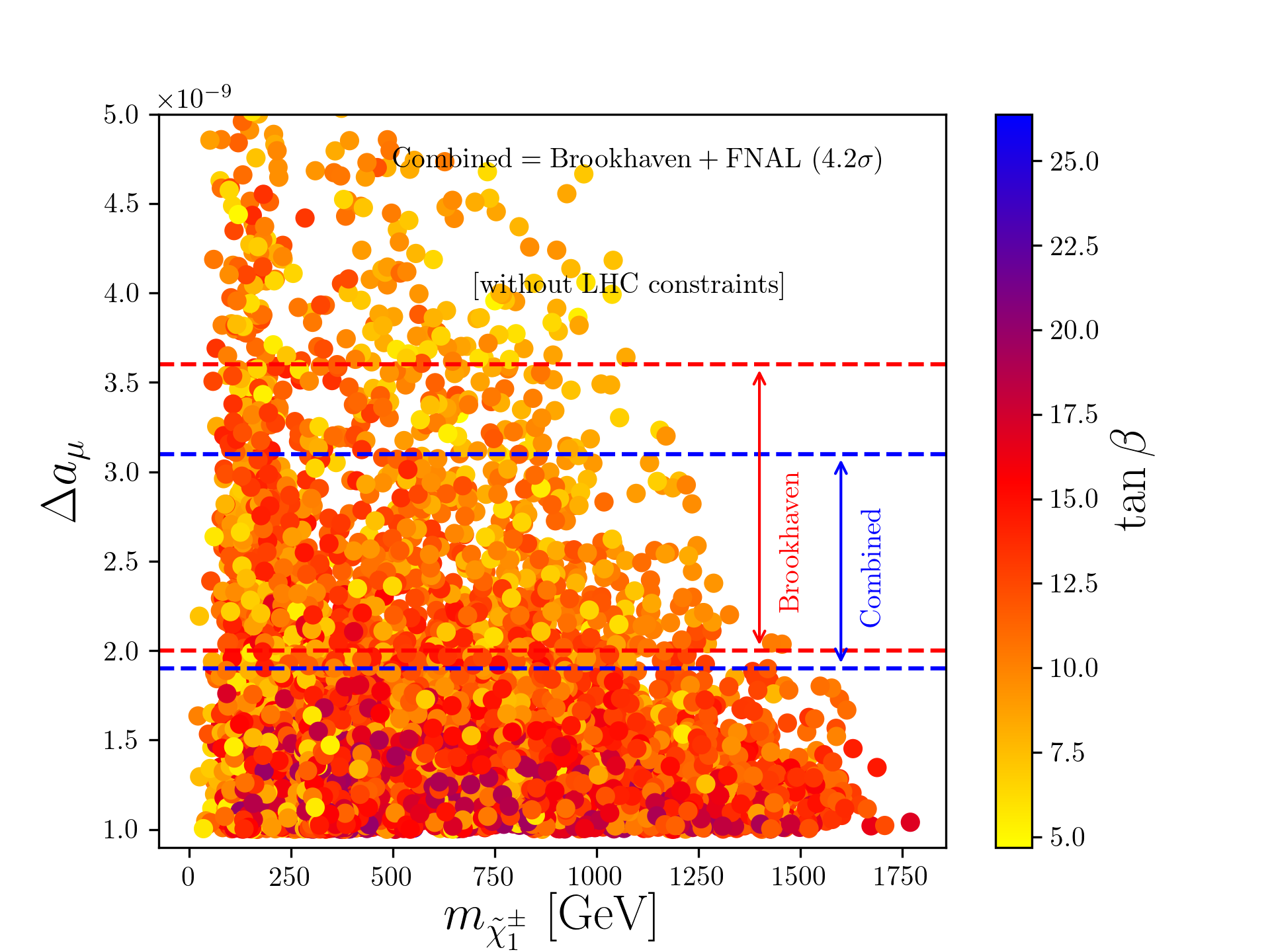}
\caption{The same scatter plots as in Figs.~\ref{fig1} (top panel) and~\ref{fig1a} but with the LHC constraints relaxed. 
We note that the dips in the model points in the smuon mass range near 600 GeV and in the chargino mass range
near 800 GeV that appeared in Figs.~\ref{fig1} (top panel) and~\ref{fig1a} are now populated.}
\label{fig1b}
\end{figure}

One final remark about Figs.~\ref{fig1} and~\ref{fig1a} is in order. 
The apparent dips in the density of models for smuons in the mass range near  600 GeV  in Figs.~\ref{fig1}
and for charginos in the mass range near 800 GeV in Figs.~\ref{fig1a}
are due to an imposition of the LHC constraints which exclude a significant number of models in this region. To exhibit this we give  the same plots in Fig.~\ref{fig1b} with the LHC constraints relaxed where one finds that the dips have disappeared. 

\begin{figure}[!htp]
\includegraphics[width=0.55\textwidth]{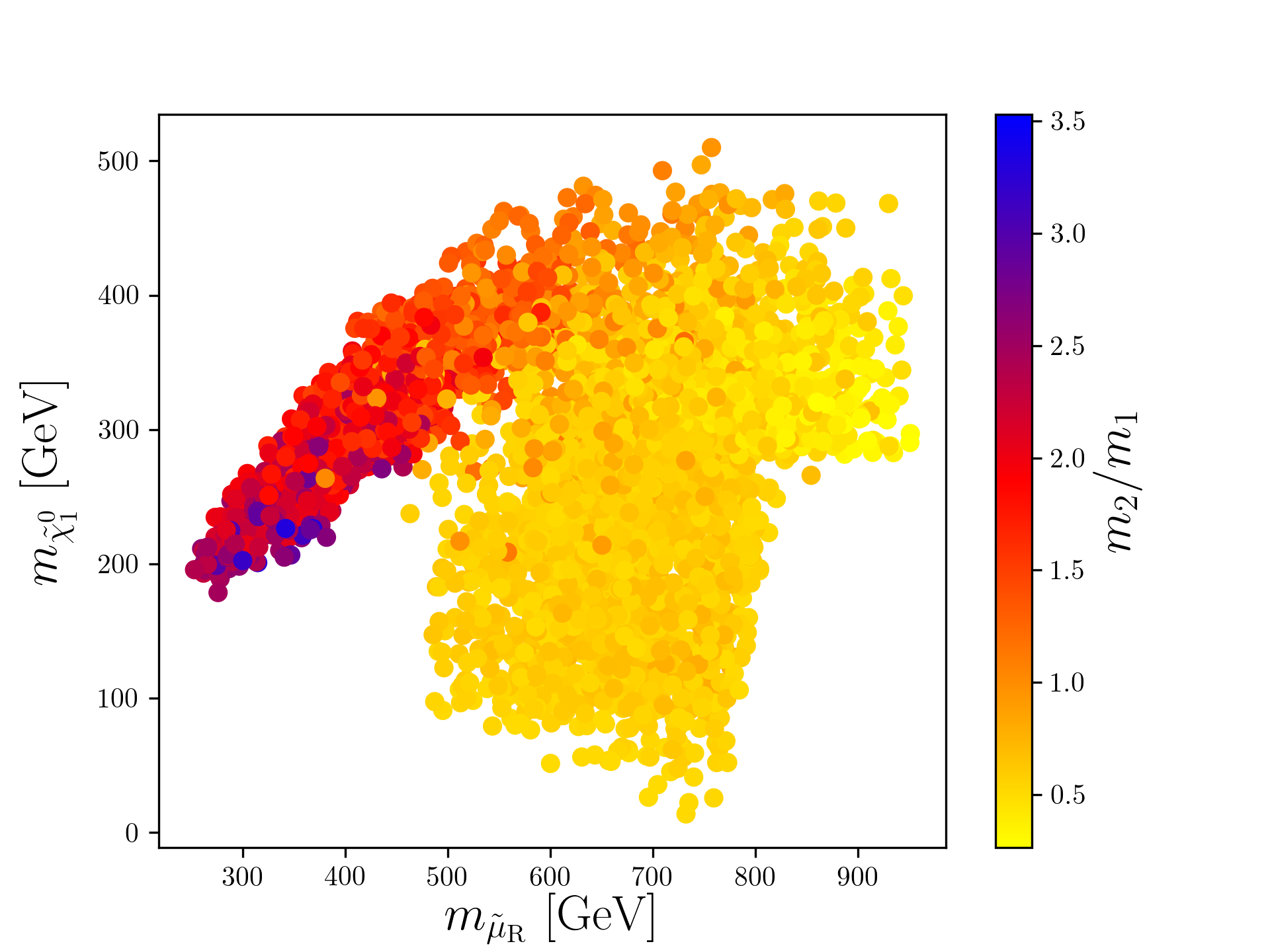}
\caption{A scatter plot in the lightest neutralino mass-right handed smuon mass plane. The color axis denotes the ratio $m_2/m_1$. Notice that the larger $m_2/m_1$ is the lighter $\tilde\mu_{\rm R}$ becomes. }
\label{fig1c}
\end{figure}

\section{3. Sparticle spectrum and dark matter constrained by \amu \label{3}}

\begin{figure}[t]
\centering
\includegraphics[width=0.48\textwidth]{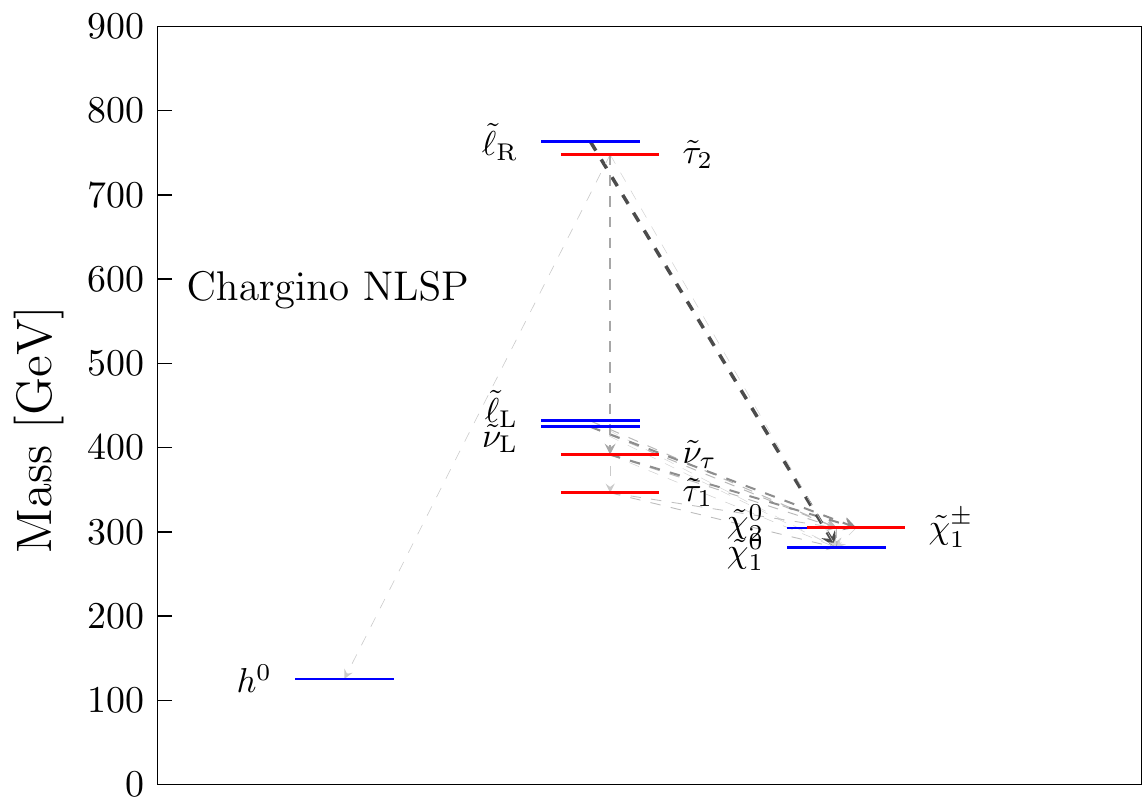}
\includegraphics[width=0.48\textwidth]{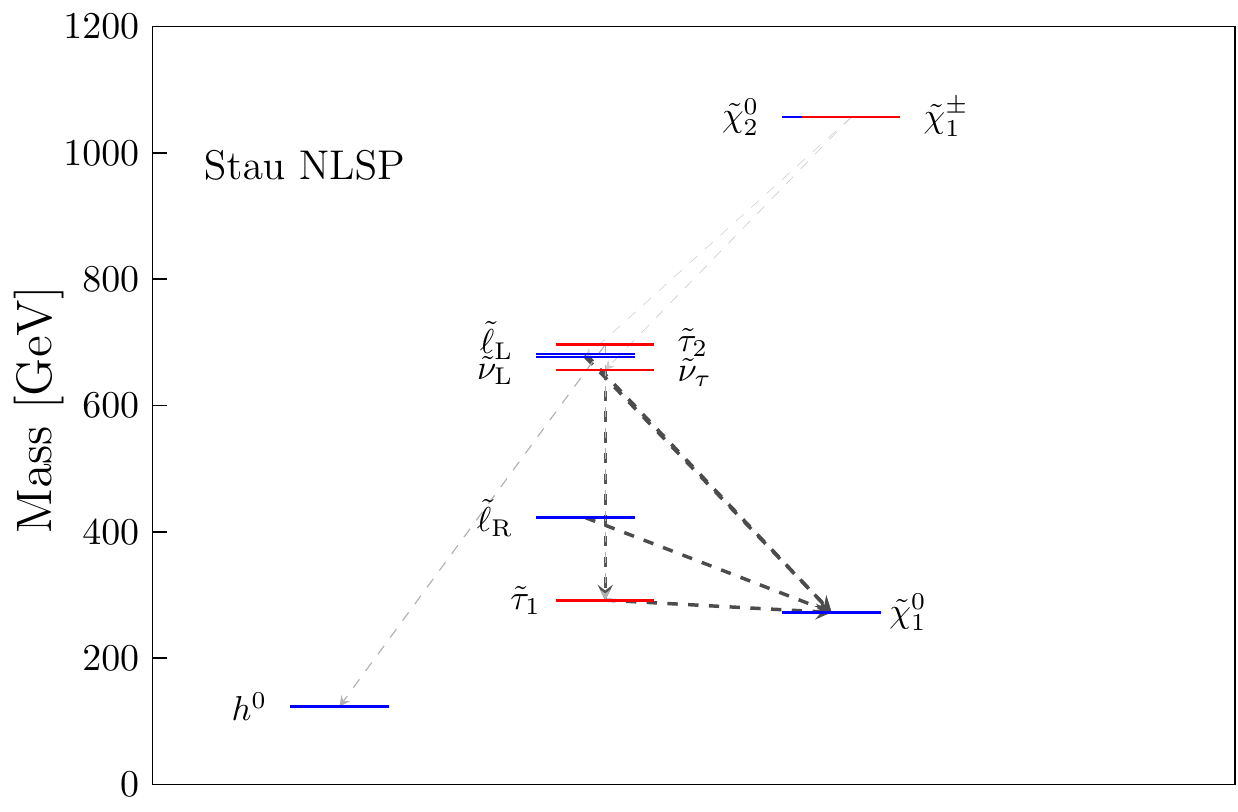}
\caption{Exhibition of the light sparticle spectrum 
in $\tilde g$SUGRA models constrained by \amu. The top panel is for the case of chargino NLSP and bottom panel is for the stau NLSP. The plots are drawn using \code{PySLHA}~\cite{Buckley:2013jua}.}
\label{figspec}
\end{figure}

As noted already,
the slepton, sneutrino and weakino  mass spectrum arising from the \amu~constraint
lie in the region of the parameter space with light and heavy particles,  where the
light particles with masses in the few hundred GeV range consisting of the neutralino, the chargino,
the smuon and muon-sneutrino produce a significant correction to $\Delta a^{\rm SUSY}_{\mu}$
while the sparticles with color, and the remaining spectrum are significantly heavy lying in the several
TeV region and do not participate in the loop corrections. 
 The mass range of the light particles is
 shown in Fig.~\ref{figspec} where the top panel exhibits an illustrative  mass range for the case when the chargino is the NLSP while the bottom panel is the case when the stau is the NLSP. The analysis shows that an smuon mass up to $\sim 1$ TeV, a chargino up to $\sim 1.5$ TeV and a neutralino up to $\sim 400$ GeV are allowed while being consistent with all constraints including \amu~and the current LHC limits. The smuon mass exhibited in Fig.~\ref{fig1} is that of the left handed  smuon. While the right handed slepton is in general heavier than the left handed one, there are regions of the parameter space where the opposite is true. The spectrum shown in Table~\ref{tab2} illustrates this phenomena,
  where for benchmarks (a)$-$(d), $m_{\tilde\ell_{\rm L}}<m_{\tilde\ell_{\rm R}}$, except for benchmark (e) where  $m_{\tilde\ell_{\rm L}}>m_{\tilde\ell_{\rm R}}$.
The reason behind this is that $m_2>m_1$ here (see Table~\ref{tab1}).
 To exhibit  this more clearly we consider the left and right smuon masses at one loop so that
  \begin{align}
m^2_{\tilde \mu_{\rm L}}&=m_0^2 + m_\mu^2 + C_1 m_1^2 + C_2 m^2_2\nonumber\\
  &+\left(-\frac{1}{2} + \sin^2\theta_W\right) M_Z^2 \cos(2\beta)\nonumber\\
 m^2_{\tilde \mu_{\rm R}}=&m_0^2 + m_\mu^2 + 4 C_1 m_1^2  - \sin^2\theta_W M_Z^2 \cos(2\beta).\nonumber
  \end{align}
  Here $C_1=\frac{3}{10}\tilde\alpha_G f_1, C_2= \frac{3}{2} \tilde \alpha_G f_2$, where $\tilde\alpha_G=\alpha_G/(4\pi)$,
$f_k(t)= t(2+b_k \tilde \alpha_G t)/(1+b_k \tilde \alpha_G t)^2$, $t=\ln(M_G^2/Q^2)$ and $(b_1,b_2)=(33/5, 1)$. 
 Thus 
 \begin{align*}
m^2_{\tilde \mu_{\rm R}} -  m^2_{\tilde \mu_{\rm L}} &=3 m_1^2C_1- m_2^2 C_2 \\
&+\left(\frac{1}{2} -2 \sin^2\theta_W\right) M_Z^2 \cos(2\beta).
  \end{align*}
The $D$-term involving $M_Z$ is relatively small for the mass ranges we are considering, and a 
numerical estimate using $\alpha_1\sim 0.016, \alpha_2=0.033, \alpha_G\sim 0.04$ and $M_G\sim 1.2 \times 10^{16}$
GeV gives $C_1\approx 0.16$ and $C_2\approx 0.23$. Thus one finds that typically the right smuon has a larger mass than the left smuon
as is seen in benchmarks (a)$-$(d) unless $m_2\gtrsim 1.24 m_1$ (with the assumed input) as seen in benchmark (e) and is supported by the
 analysis of  Fig.~\ref{fig1c}.

\begin{figure}[t]
\centering
\vspace{1mm}
\includegraphics[width=0.5\textwidth]{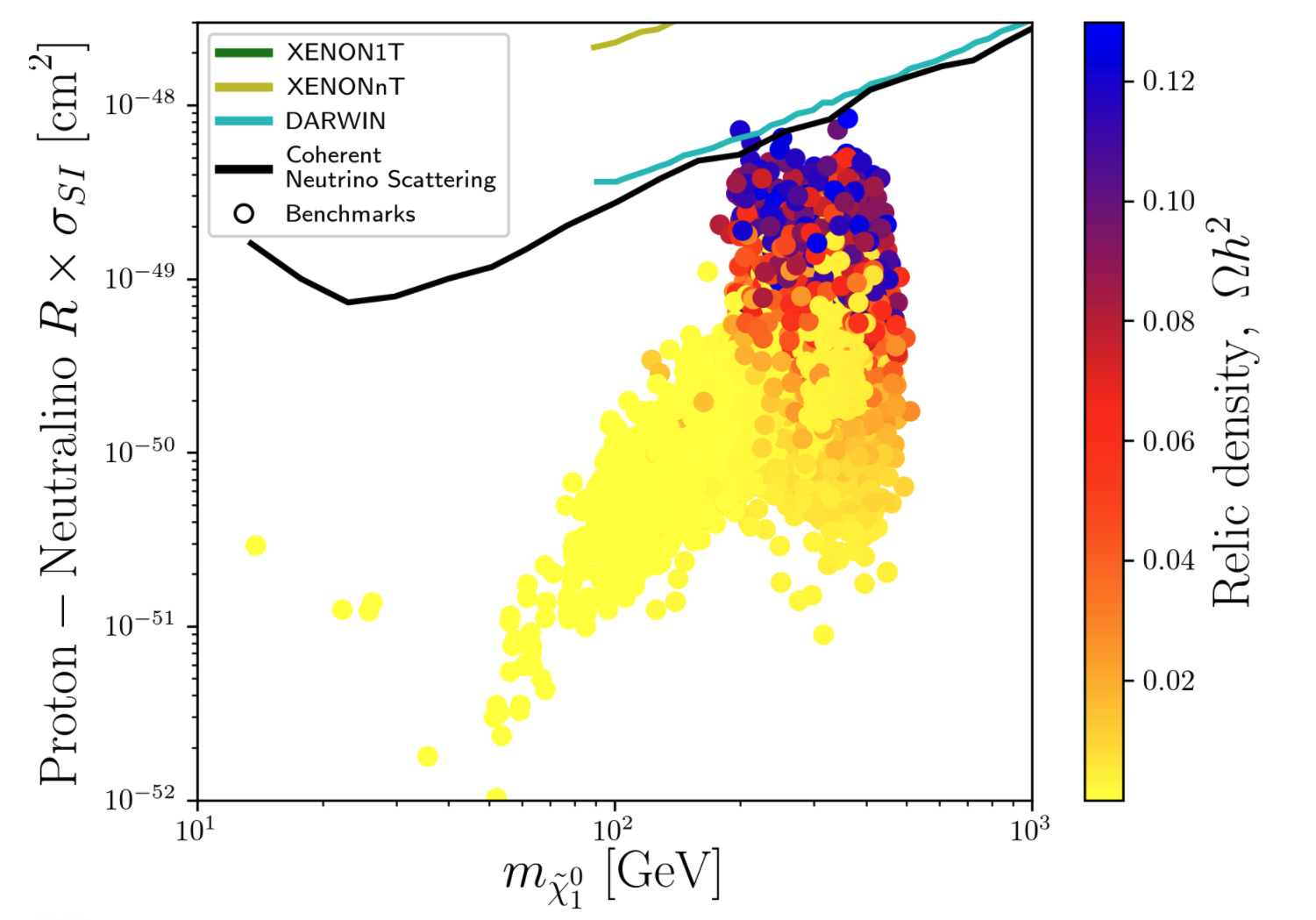}
\caption{Spin-independent proton-neutralino cross section $\sigma_{\rm SI}$ as a function of the 
neutrino mass constrained by  \amu.  Here $R=\Omega h^2/(\Omega h^2)_{\rm Planck}$ where $(\Omega h^2)_{\rm Planck}\sim 0.12$. }
\label{figdark}
\end{figure}

 \amu~also puts significant constraints on the spin-independent neutralino-nucleon
cross section $\sigma_{\rm SI}$.  As shown in Fig.~\ref{figdark}, one finds that 
some of the models have
$\sigma_{\rm SI}$  within  reach of DARWIN~\cite{Macolino:2020uqq}
and  are thus discoverable. However, most of the allowed parameter space consistent with the \amu~constraint lies below the neutrino floor. The smallness of the $\sigma_{\rm SI}$ is a direct consequence of the fact that the neutralino is mostly
a bino with only a small wino content. 

\section{4. Implications of \amu~for discovering SUSY at HL-LHC and HE-LHC \label{4}}

 The light sleptons and sneutrinos appearing in  Table~\ref{tab2} could be pair produced  
  in proton-proton collisions at 14 TeV (HL-LHC) and 27 TeV (HE-LHC). Another important production mode is  the associated production of a slepton with an sneutrino which can have a significantly larger cross section than 
  those of light sleptons and of light sneutrinos.
  The production cross-sections are computed at aNNLO+NNLL accuracy using~\code{Resummino-3.0}~\cite{Debove:2011xj,Fuks:2013vua} and the five-flavor NNPDF23NLO PDF set and given in Tables~\ref{tab3} and~\ref{tab4}.
Slepton (selectron and smuon) and sneutrino production constitute a difficult signal region to look for at the LHC owing to their small production cross section and the decay topology resembling the SM backgrounds. The muon $g-2$ prefers smuons (and sneutrinos) with mass less than 1 TeV as one can see from Fig.~\ref{fig1} and their direct detection at the LHC is of importance especially after the recent $(g-2)_{\mu}$ results from Fermilab. 
Our signal consists of smuon (muon sneutrino) and selectron (electron sneutrino) pair production as well as slepton associated production with a sneutrino which decay to light leptons (electrons and muons) and a neutralino. The final states which make up our signal region involve two same flavor and opposite sign (SFOS) leptons with missing transverse energy (MET). We consider two main signal regions where for one signal region we require exactly one isolated jet which can be used to trigger on especially in an initial state radiation (ISR)-assisted topology when the MET is small, and for another signal region we require at least two jets targeting benchmarks with jetty final states. We call the former signal region SR-2$\ell$1j and the latter SR-2$\ell$2j. For such final states, the dominant SM backgrounds are from diboson production, $Z/\gamma+$jets, dilepton production from off-shell vector bosons ($V^*\rightarrow\ell\ell$), $t\bar{t}$ and $t+W/Z$. The subdominant backgrounds are Higgs production via gluon fusion ($ggF$ H) and vector boson fusion (VBF).   The simulation of the signal and background events is performed at LO with \code{MadGraph5\_aMC@NLO-3.1.0}~\cite{Alwall:2014hca} interfaced to \code{LHAPDF}~\cite{Buckley:2014ana} using the NNPDF30LO PDF set. Up to two hard jets are added at generator level. The parton level events are passed to \code{PYTHIA8}~\cite{Sjostrand:2014zea} for showering and hadronization using a five-flavor matching scheme in order to avoid double counting of jets. For the signal events, the matching/merging scale is set at one-fourth the mass of the pair produced sleptons or sneutrinos. Additional jets from ISR and FSR are added to the signal and background events. Jets are clustered with \code{FastJet}~\cite{Cacciari:2011ma} using the anti-$k_t$ algorithm~\cite{Cacciari:2008gp} with jet radius $R=0.4$. \code{DELPHES-3.4.2}~\cite{deFavereau:2013fsa} is then employed for detector simulation and event reconstruction using the HL-LHC and HE-LHC card. The SM backgrounds are scaled to their relevant NLO cross sections while aNNLO+NNLL cross sections are used for the signal events. 

\begin{table}[t]
\caption{\label{tab3} The aNNLO+NNLL pair production cross-sections, in fb, of sleptons at $\sqrt{s}=14$ TeV and at $\sqrt{s}=27$ TeV for benchmarks (a)$-$(e) of Table~\ref{tab1}.} 
\begin{ruledtabular}
\begin{tabular}{ccc|cc}
Model & \multicolumn{2}{c|}{$\sigma(pp\rightarrow \tilde e_L\,\tilde e_L~[\tilde e_R\,\tilde e_R])$}& \multicolumn{2}{c}{$\sigma(pp\rightarrow \tilde\mu_L\,\tilde\mu_L~[\tilde\mu_R\,\tilde\mu_R])$}  \\
\hline
&14 TeV & 27 TeV & 14 TeV & 27 TeV\\
\hline \rule{0pt}{3ex}
\!\!(a) & 4.41 [0.159] & 14.0 [0.684] & 4.42 [0.159] & 14.0 [0.684]   \\
 (b) & 1.39 [0.028] & 5.07 [0.168] & 1.40 [0.028] & 5.09 [0.168] \\
  (c) & 0.58 [0.029] & 2.38 [0.171] & 0.58 [0.029] & 2.39 [0.171]  \\ 
  (d) & 4.90 [0.328] & 15.4 [1.27] & 4.91 [0.328] & 15.4 [1.27]\\ 
  (e) & 0.095 [0.493] & 0.54 [1.79] & 0.096 [0.495] & 0.54 [1.80] 
\end{tabular}
\end{ruledtabular}
\end{table}

The discrimination between the signal and background events is done with the help of a deep neural network (DNN) as part of the `Toolkit for Multivariate Analysis' (TMVA)~\cite{Speckmayer:2010zz} framework within \code{ROOT6}~\cite{Antcheva:2011zz}. To train the signal and background events, we use a set of discriminating variables:
\begin{enumerate}

\item $E^{\rm miss}_T$: the missing transverse energy in the event. It is usually high for the signal due to the presence of neutralinos.

\item The transverse momentum of the leading non-b tagged jets, $p_T(j_1)$. Rejecting b-tagged jets reduces the $t\bar{t}$ background.

\item The transverse momentum of the leading and subleading leptons (electron or muon), $p_T(\ell_1)$ and $p_T(\ell_2)$, respectively. 

\item The total transverse momentum of all the ISR jets in an event, $p_T^{\rm ISR}$.

\item $M_{\rm T2}$, the stransverse mass~\cite{Lester:1999tx, Barr:2003rg, Lester:2014yga} of the leading and subleading leptons
\begin{equation}
    M_{\rm T2}=\min\left[\max\left(m_{\rm T}(\mathbf{p}_{\rm T}^{\ell_1},\mathbf{q}_{\rm T}),
    m_{\rm T}(\mathbf{p}_{\rm T}^{\ell_2},\,\mathbf{p}_{\rm T}^{\text{miss}}-
    \mathbf{q}_{\rm T})\right)\right],
    \label{mt2}
\end{equation}
where $\mathbf{q}_{\rm T}$ is an arbitrary vector chosen to find the appropriate minimum and the transverse mass $m_T$ is given by 
\begin{equation}
    m_{\rm T}(\mathbf{p}_{\rm T1},\mathbf{p}_{\rm T2})=
    \sqrt{2(p_{\rm T1}\,p_{\rm T2}-\mathbf{p}_{\rm T1}\cdot\mathbf{p}_{\rm T2})}.
\end{equation}  

\item The quantity $M^{\rm min}_{\rm T}$ defined as $M^{\rm min}_{\rm T}=\text{min}[m_{\rm T}(\textbf{p}_{\rm T}^{\ell_1},\textbf{p}^{\rm miss}_{\rm T}),m_{\rm T}(\textbf{p}_{\rm T}^{\ell_2},\textbf{p}^{\rm miss}_{\rm T})]$. 
The variables $M_{\rm T2}$ and $M^{\rm min}_{\rm T}$ are effective when dealing with large MET in the final state.

\item The dilepton invariant mass, $m_{\ell\ell}$, helps in rejecting the diboson background with a peak near the $Z$ boson mass which can be done by requiring $m_{\ell\ell}>100$ GeV.

\item The opening angle between the MET system and the dilepton system, $\Delta\phi(\textbf{p}_{\rm T}^{\ell},\textbf{p}^{\rm miss}_{\rm T})$, where $\textbf{p}_{\rm T}^{\ell}=\textbf{p}_{\rm T}^{\ell_1}+\textbf{p}_{\rm T}^{\ell_2}$. 

\item The smallest opening angle between the first three leading jets in an event and the MET system, $\Delta\phi_{\rm min}(\textbf{p}_{\rm T}(j_i),\textbf{p}^{\rm miss}_{\rm T})$, where $i=1,2,3$.

\end{enumerate} 

\begin{figure}[!htp]
\centering
\includegraphics[width=0.5\textwidth]{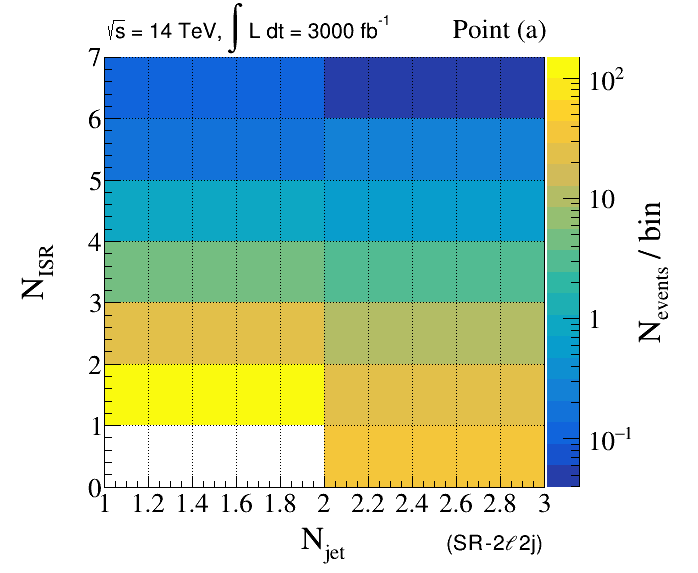}
\caption{A two-dimensional plot in the number of ISR jets ($N_{\rm ISR}$) versus non-ISR jets ($N_{\rm jet}$). }
\label{isr}
\end{figure}

It is worth mentioning how jets are classified as either coming from an ISR or from the decay of the SUSY system. After reconstructing the momentum of the dilepton system, we determine the angle between the dilepton system and each non-b-tagged jet in the event, i.e., $\Delta\phi(p_T(j_i),p_T^{\ell})$. If an event has exactly two jets with leading and subleading transverse momenta, $p_T(j_1)$ and $p_T(j_2)$, respectively, then both are tagged as non-ISR if $\Delta\phi(p_T(j_1),p_T^{\ell})<\Delta\phi(p_T(j_2),p_T^{\ell})$. However, if $\Delta\phi(p_T(j_1),p_T^{\ell})>\Delta\phi(p_T(j_2),p_T^{\ell})$, then the subleading jet is tagged as non-ISR and the leading one will be an ISR jet.  If an event has more than two jets, then we select \textit{up to two} jets that are closest to the dilepton system and tag them as non-ISR (possible jets arising from the decay of the SUSY system) and the rest are classified as ISR jets. Fig.~\ref{isr} shows a 2D plot in the number of jets tagged as ISR ($y$ axis) versus the number of non-ISR jets ($x$ axis). One can see that the largest number of events correspond to the case of one ISR and one non-ISR jet per event. Moreover, one can get as many as six ISR jets in an event but with a low event count while a larger number of events have no ISR jets. 

\begin{widetext}

\begin{figure}[!htp]
\centering
\includegraphics[width=0.32\textwidth]{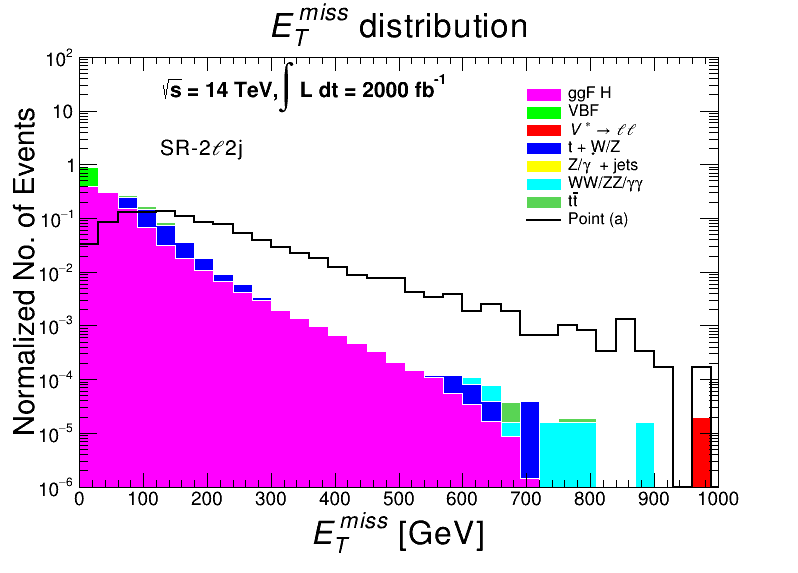}
\includegraphics[width=0.32\textwidth]{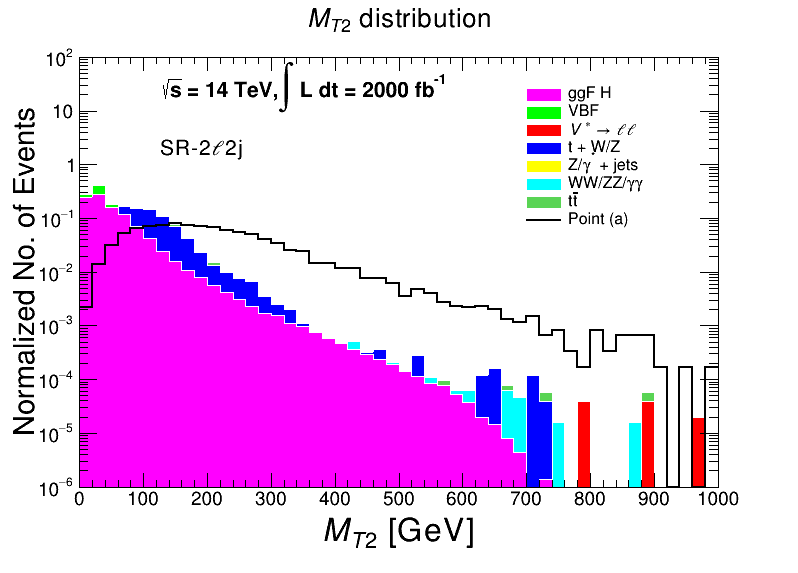}
\includegraphics[width=0.32\textwidth]{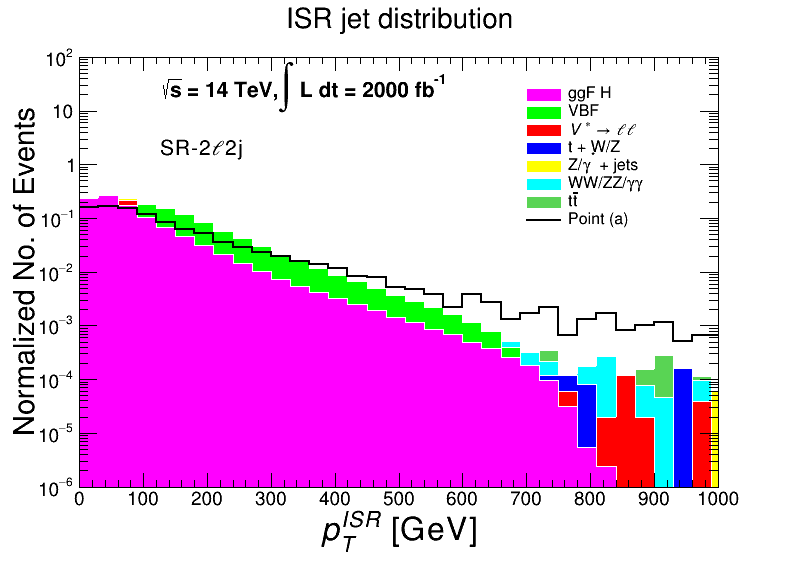}\\
\includegraphics[width=0.32\textwidth]{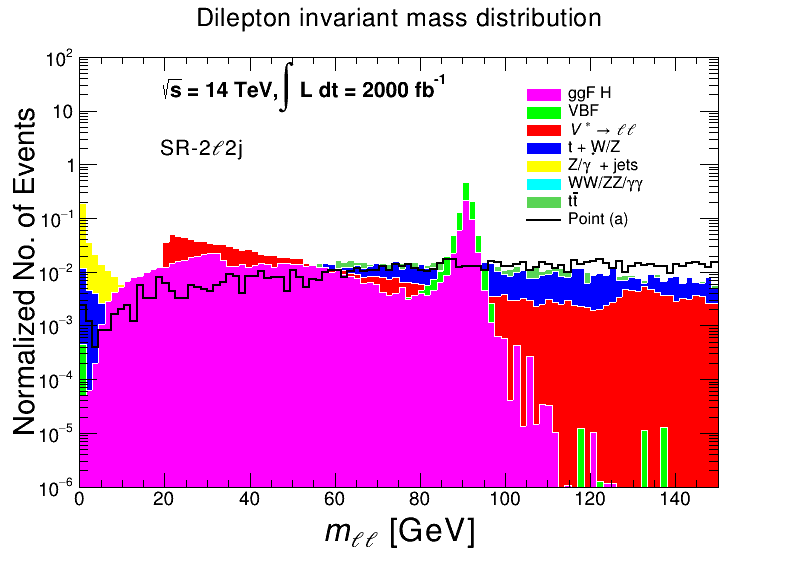}
\includegraphics[width=0.32\textwidth]{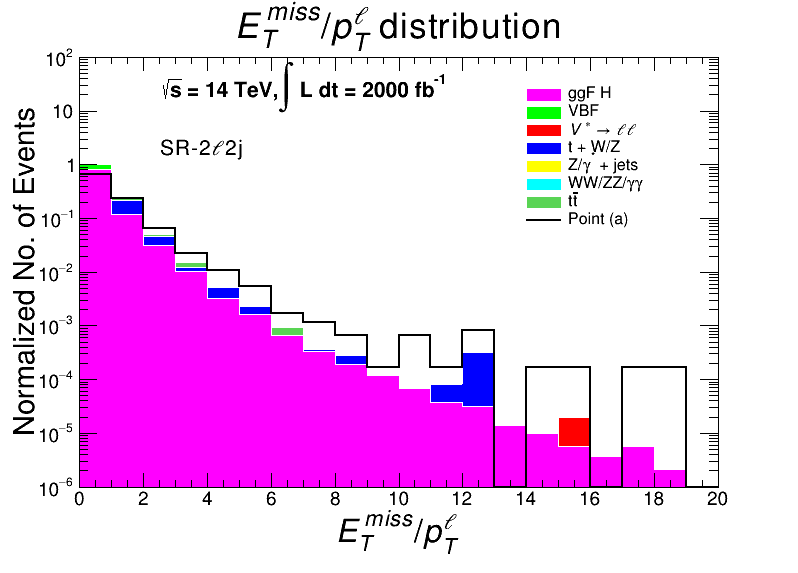}
\includegraphics[width=0.32\textwidth]{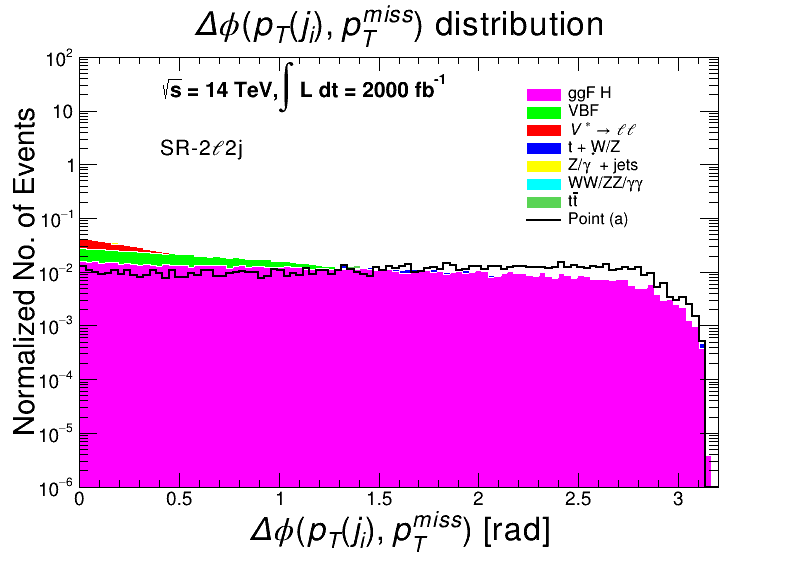}
\caption{A sample of the discriminating variables using by the DNN for training and testing. The distributions are in the normalized number of events scaled by a specific integrated luminosity to show the discriminating power of each variable.}
\label{figk}
\end{figure}

\end{widetext}

Fig.~\ref{figk} shows normalized distributions in six of the discriminating variables which are used by the DNN for training.
 Before the events are fed into a DNN, a set of preselection criteria is applied to the signal and background. The leading and subleading leptons must have a transverse momenta $p_T>15$ GeV for electrons and $p_T>10$ GeV for muons with $|\eta|<2.5$. Each event in SR-2$\ell$1j should contain exactly one non-b-tagged jet while in SR-2$\ell$2j at least two non-b-tagged jets are required with the leading $p_T>20$ GeV in the $|\eta|<2.4$ region and $E^{\rm miss}_T>100$ GeV. The preselection criteria are summarized in Table~\ref{cuts}.

\subsection{Slepton pair production} 

We begin the analysis with the first production mode which is slepton pair production. Table~\ref{tab3} shows the pair production cross sections of left handed and right handed sleptons for the benchmarks of Table~\ref{tab1} at 14 TeV and 27 TeV. Notice that the contribution from right handed sleptons is small compared to the left handed ones except for benchmark (e) where $m_{\tilde\ell_{\rm R}}<m_{\tilde\ell_{\rm L}}$. In this model point, the right handed slepton has a mass $\sim 400$ GeV which is comparable to the left handed slepton of benchmark (b). However, the cross section is smaller as one expects. Those benchmarks represent a contrast between high scale models and simplified models considered in LHC analyses. Thus, in ATLAS and CMS analyses, left and right handed sleptons are considered to be of the same mass and are excluded on an equal footing. In our case, however, we must consider the two particles separately. Of course, if a specific benchmark has a left (right) handed slepton which is excluded by experiment then this would eliminate the entire benchmark regardless of the mass of its right (left) handed counterpart. In this analysis we focus on the left handed sleptons knowing that the right handed ones have less significant contribution but we make sure that the right handed sleptons are not excluded as this would entirely eliminate the benchmark under study. An exception to this situation is benchmark (e), where despite having a small $\tilde\ell_{\rm R}\tilde\ell_{\rm R}$ production cross section compared to its left handed counterpart of the same mass, the branching ratio of $\tilde\ell_{\rm R}$ to $\ell\tilde\chi^0_1$ is unity (see Table~\ref{tab2a}) which makes $\sigma\times\text{BR}$ significant. Hence in this benchmark, we simulate both the left and right handed sleptons. Another aspect of high scale models which differentiates our analysis from that of LHC concerns 
 the branching ratios of slepton and sneutrino decays. 
 The relevant  branching ratios are given in Table~\ref{tab2a} and unlike LHC analyses which consider a unit branching fraction to leptons and neutralinos, our benchmarks have a more diverse decay topology.
Next, we present the results from the two signal regions SR-2$\ell$2j and SR-2$\ell$1j defined earlier.      

\paragraph{\textbf{The signal region SR-2$\ell$2j:}}  
 
\begin{figure}[t]
\centering
\includegraphics[width=0.48\textwidth]{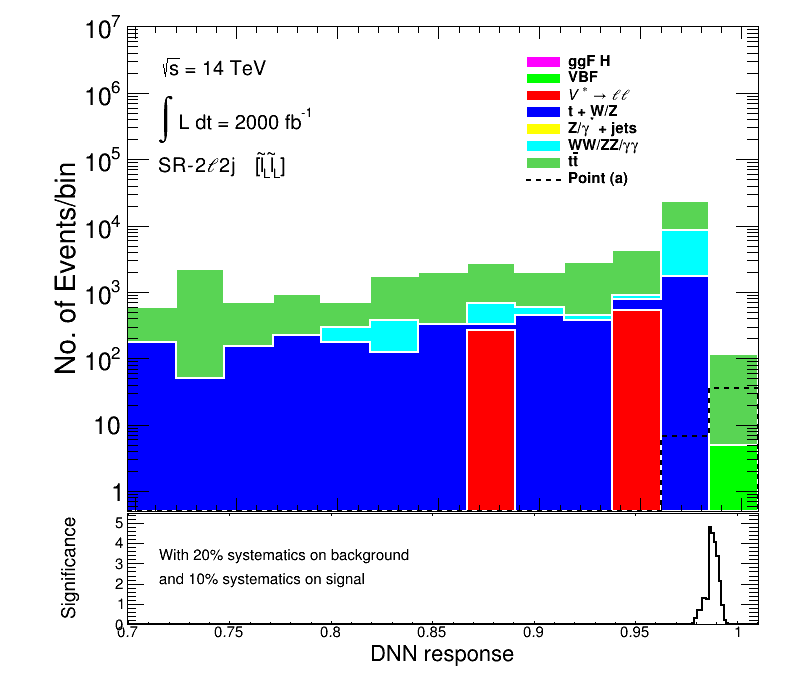}
\includegraphics[width=0.48\textwidth]{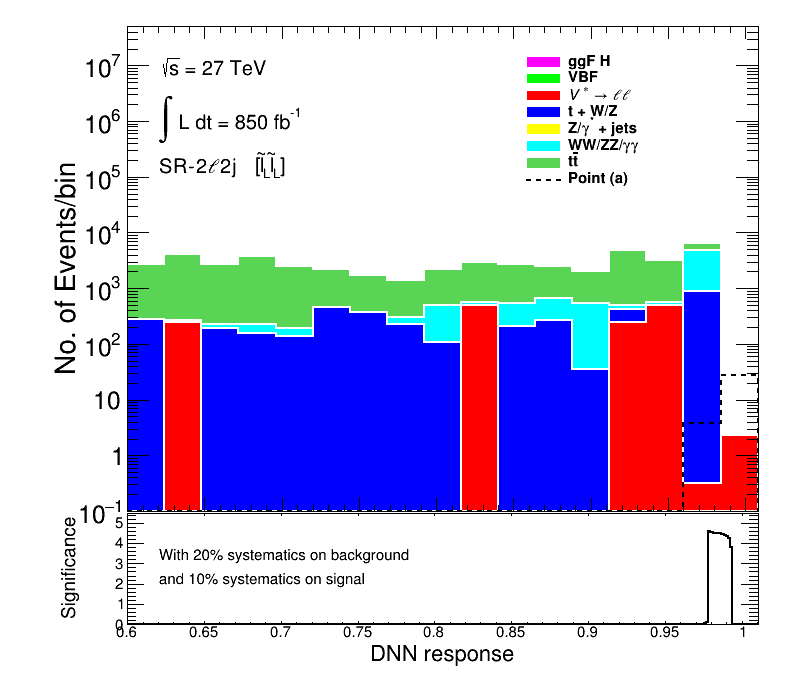}
\caption{Distributions in the DNN response for the signal (dashed histogram) and background (colored histograms) events pertaining to benchmark (a) at 14 TeV (top panel) and 27 TeV (bottom panel) for the signal region SR-$2\ell 2$j in the slepton pair production channel. The bottom pad of each panel shows the significance as defined by Eq.~(\ref{significance}) as a function of the cut on the `DNN response' variable. The binning for the significance distribution is finer to clearly show the rise and fall of the significance.  }
\label{figdnn1}
\end{figure}

To train and test the signal ($S$) and background ($B$) events that have passed the preselection criteria, a four-layer DNN uses two statistically independent sets of signal and background events. The training phase employs the above set of variables to create a new powerful kinematic variable called the `DNN response' which can be used as a discriminant to reject events thus maximizing the $S/\sqrt{S+B}$ ratio. Fig.~\ref{figdnn1}  shows distributions in the DNN response for benchmark (a) at 14 TeV and 27 TeV. The DNN has successfully separated the signal events which can be seen peaking near 1 while the SM background is more concentrated at values less than 1. The cut on the DNN response is aided by a series of analysis cuts using some of the variables described above. A summary of the preselection criteria and the analysis cuts is given in Table~\ref{cuts}.  A minimum cut on $\mathrm{DNN~response}>0.95$ removes most of the background and benchmarks (a)$-$(e) become discoverable at both HL-LHC and HE-LHC. The estimated integrated luminosities for discovery are shown in the last two columns of Table~\ref{cuts}. It is worth mentioning that benchmark (e) becomes discoverable at HL-LHC only when the contribution from right handed sleptons is included.   We note that the cuts need to be customized when studying HE-LHC as compared to HL-LHC.

\begin{widetext}

\begin{table}[H]
\caption{The analysis uses cuts on a set of kinematic variables at 14 TeV (27 TeV) grouped by the benchmarks of Table~\ref{tab1} in two signal regions SR-$2\ell1$j and SR-$2\ell2$j. We note  that with the exception of $m_{\ell\ell}$ harder cuts are applied at 27 TeV. Entries with a dash (-) imply  that no requirement on the variable is considered.  Shown at the bottom of the table are the integrated luminosities needed for discovery at 14 TeV and 27 TeV. Also shown are the preselection criteria used.}
\label{cuts}
\begin{ruledtabular}
\begin{tabular}{|c|ccc|ccc|}
\multirow{2}{*}{Observable}  & (a), (b), (d) & (c) & (e)  & (a), (b), (d) & (c) & (e) \\
\cline{2-7}\rule{0pt}{3ex}
 & \multicolumn{3}{c|}{Preselection criteria (SR-$2\ell2$j)} & \multicolumn{3}{c|}{Preselection criteria (SR-$2\ell1$j)} \\
 \hline \rule{0pt}{3ex}
 $N_{\ell}$ (SFOS) & \multicolumn{3}{c|}{$2$} & \multicolumn{3}{c|}{$2$} \\
$N_{\rm jets}^{\rm non-b-tagged}$ & \multicolumn{3}{c|}{$\geq 2$} & \multicolumn{3}{c|}{$1$}  \\
$p_T(j_1)$ [GeV] & \multicolumn{3}{c|}{$>20$} & \multicolumn{3}{c|}{$>20$} \\
$p_T(\ell_1)$ (electron, muon) [GeV] & \multicolumn{3}{c|}{$>15$, $>10$} & \multicolumn{3}{c|}{$>15$, $>10$} \\
$E^{\rm miss}_T$ [GeV] & \multicolumn{3}{c|}{$>100$} & \multicolumn{3}{c|}{$>100$}\\
\cline{2-7}\rule{0pt}{3ex}
 & \multicolumn{3}{c|}{Analysis cuts} & \multicolumn{3}{c|}{Analysis cuts} \\
\cline{2-7}\rule{0pt}{3ex}
\!\! $m_{\ell\ell}~\text{[GeV]} >$ & 130 & 150  & 150 (110) & 130 (240) & 200 (150)  & 150 (110)   \\
 $E^{\rm miss}_T/\textbf{p}^{\ell}_{\rm T}>$ & 0.5 (2.8)  &  -  & -  & 1.0 (1.5)  &  -  & -   \\
$\Delta\phi_{\rm min}(\textbf{p}_{\rm T}(j_i),\textbf{p}^{\rm miss}_{\rm T})~\text{[rad]} >$ & -  & 0.85 (1.5)  &  -  & -  & 0.80 (1.5)  &  - \\
$p_T(\ell_2)~\text{[GeV]} >$ & -  &  -  & 190 (370) & -  &  -  & 190 (300)   \\
$M_{T2}~\text{[GeV]} >$ & 120 (140)  &  120  & 200 (300) & -   &  120  & 200 (300)  \\
DNN response $>$   & 0.95  & 0.95 & 0.95 & 0.95  & 0.95 & 0.95 \\
  \hline
  $\mathcal{L}$ at 14 TeV [fb$^{-1}$] & 1629, 1559, 1371 & 664 & 1292 & 426, 853, 478 & 2742 & 923 \\
$\mathcal{L}$ at 27 TeV [fb$^{-1}$] & 716, 1432, 535 & 314 & 827 & 306, 387, 347 & 830 & 572 \\
\end{tabular}
\end{ruledtabular}
\end{table}

\end{widetext}

\begin{figure}[!htp]
\centering
\includegraphics[width=0.48\textwidth]{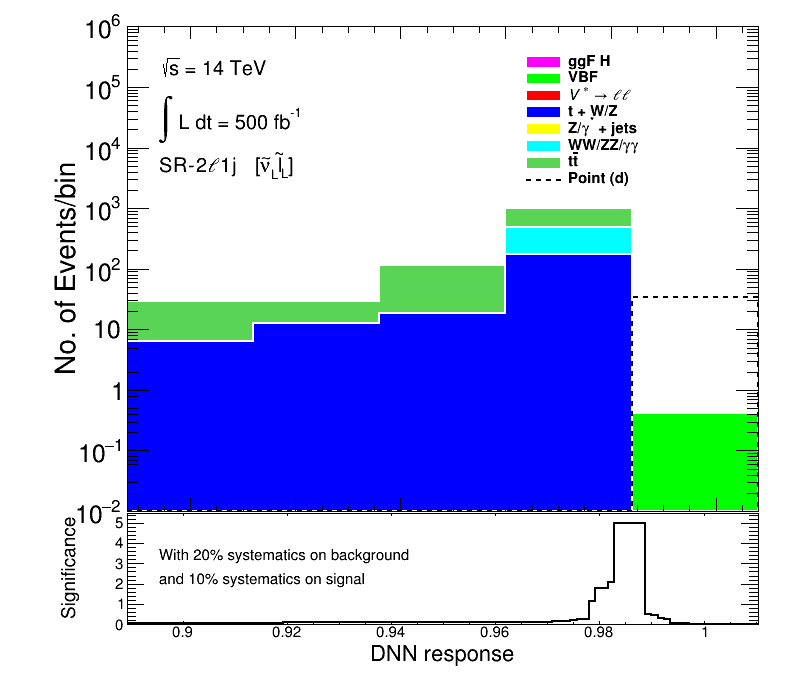}
\includegraphics[width=0.48\textwidth]{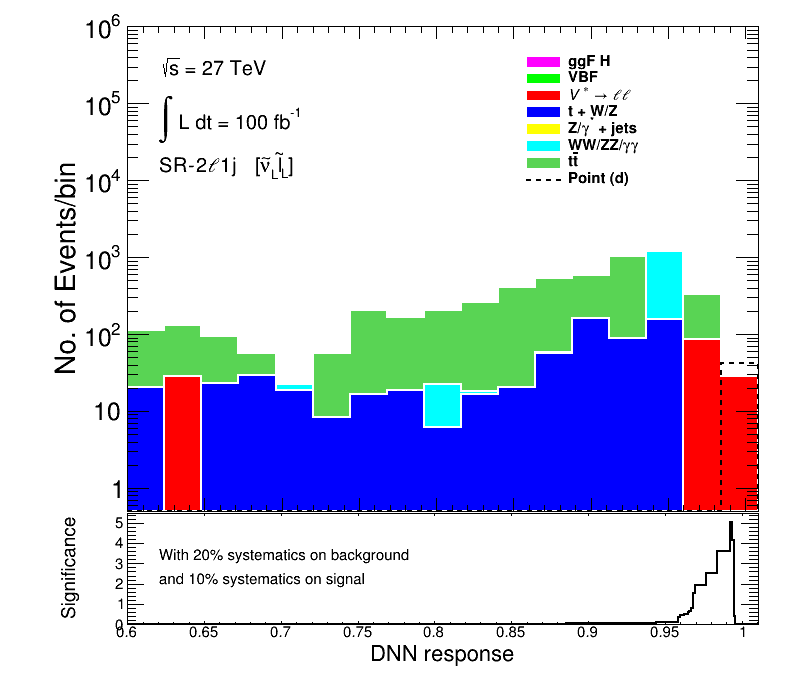}
\caption{Distributions in the DNN response for the signal (dashed histogram) and background (colored histograms) events pertaining to benchmark (d) at 14 TeV (top panel) and 27 TeV (bottom panel) for the signal region SR-$2\ell 1$j in the slepton associated production channel. The bottom pad of each panel shows the significance as defined by Eq.~(\ref{significance}) as a function of the cut on the `DNN response' variable. The binning for the significance distribution is finer to clearly show the rise and fall of the significance.}
\label{figdnn2}
\end{figure}

\begin{figure}[t]
\centering
\includegraphics[width=0.5\textwidth]{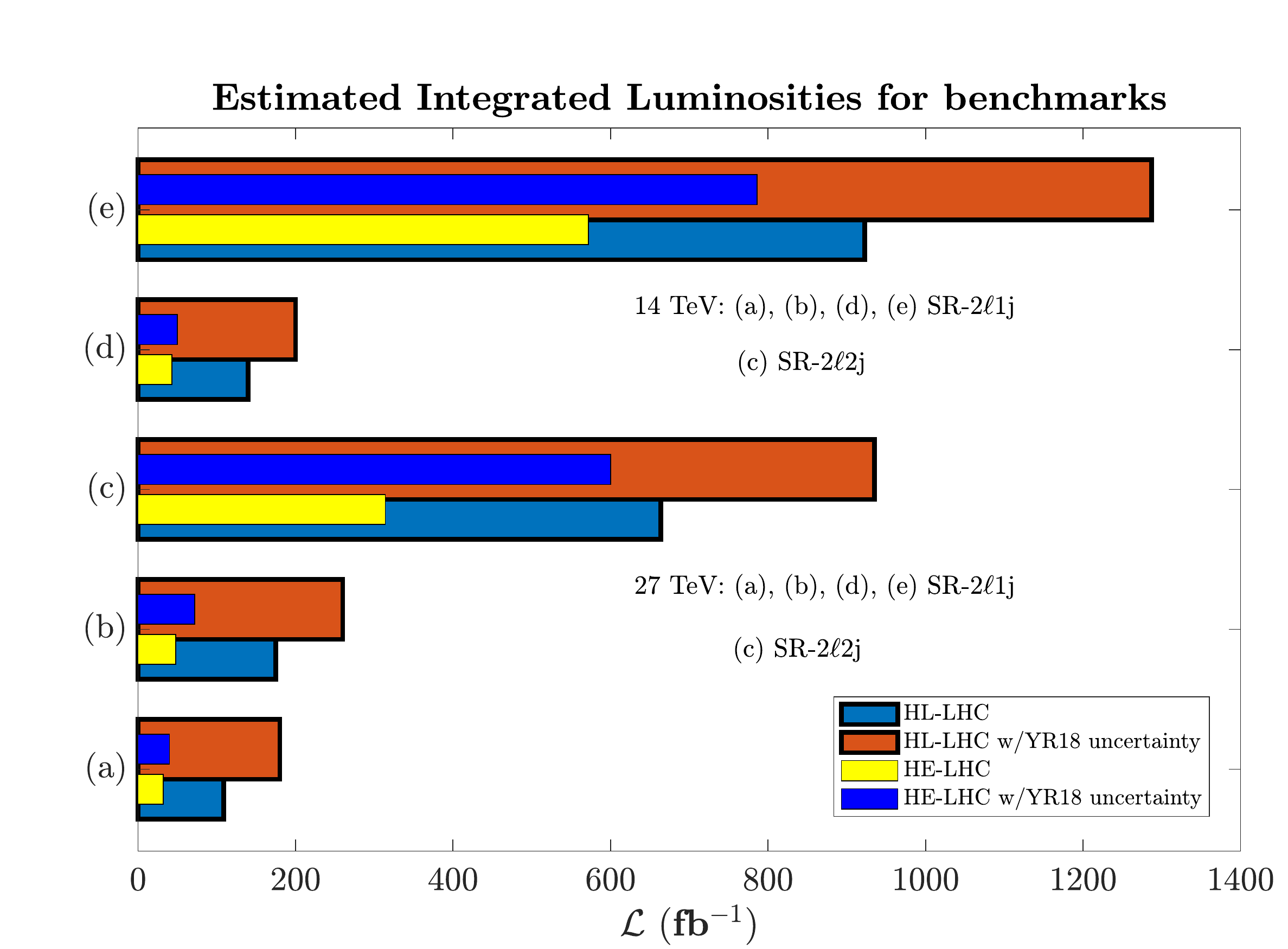}
\caption{The integrated luminosities needed for discovery of 
SUSY at HL-LHC and HE-LHC assuming that \amu~arises from SUSY loops. The signal regions and production modes shown are the ones giving the highest sensitivity for discovery. Also shown are the integrated luminosities after including the `YR18' uncertainties on the signal and background. }
\label{figlumi}
\end{figure}

\paragraph{\textbf{The signal region SR-2$\ell$1j:}}  

In this signal region, we require only one non-b-tagged jet which has the potential of offering a greater sensitivity~\cite{Aad:2019vnb}. In this signal region, we do not differentiate between ISR and non-ISR jets. Therefore the variable $p_T^{\rm ISR}$ is not used here. Using the same DNN training and testing technique discussed in the preceding 
analysis, we construct the `DNN response' variable and apply the selection criteria specific to this signal region as shown in Table~\ref{cuts}. We then estimate the integrated luminosity required for a $5\sigma$ discovery for each benchmark. The results are shown in the last two columns of Table~\ref{cuts}. 
We note that that the single jet signal region provides a greater sensitivity for detection relative to the two jet signal
  for benchmarks (a), (b) and (d). 
On the other hand the two-jet signal region shows a better detection sensitivity for benchmark (c) than the 
  single jet signal region. The reason is that benchmark (c) has
   a stau  which is the NLSP. So the decay channels $\tilde\ell_{\rm L}\to\tilde\chi^0_2\ell\to\tilde\tau\tau\ell\to\tau\tau\ell\tilde\chi^0_1$ and $\tilde\ell_{\rm L}\to\tilde\chi_1^{-}\nu_{\ell}\to\tau\nu_\tau\nu_{\ell}\tilde\chi^0_1$ render a tau-enriched final state. Since taus can form jets, then requiring at least two jets in SR-2$\ell$2j does lead to a better sensitivity than SR-2$\ell$1j. 

\subsection{Sneutrino pair production}

According to Table~\ref{tab2a}, the decay channel $\tilde\nu_{\rm L}\to\tilde\chi^+_1\ell^-$ has a significant branching ratio for benchmarks (a), (b) and (d) which correspond to the case of a chargino NLSP. Since the chargino and the LSP are nearly degenerate, the decay products cannot be discerned and therefore would contribute to the total MET.
  In this case, the final states will be identical to the slepton pair production mode discussed in the previous section. The sneutrino pair production cross section at aNNLO+NNLL is given in Table~\ref{tab4} for benchmarks (a), (b) and (d).  Since we have already shown that the signal region SR-$2\ell1$j provides a better sensitivity for these benchmarks, we will use it again to estimated the required integrated luminosity for discovery at 14 TeV and 27 TeV. The results are shown in Table~\ref{tab5}. In comparison to slepton pair production, the sneutrino pair production mode fairs better at the chances of discovering SUSY with benchmark (d) requiring only 317 fb$^{-1}$ of integrated luminosity which should become available in the next round of data taking at the LHC. That is contrasted with 1371 fb$^{-1}$ needed in SR-$2\ell2$j and 478 fb$^{-1}$ in SR-$2\ell1$j for the slepton pair production mode as shown in Table~\ref{cuts}.

\begin{table}[t]
\caption{\label{tab4} The aNNLO+NNLL pair production cross-sections, in fb, of sneutrinos and the slepton associated production at $\sqrt{s}=14$ TeV and at $\sqrt{s}=27$ TeV for benchmarks (a), (b) and (d) of Table~\ref{tab1}.} 
\begin{ruledtabular}
\begin{tabular}{ccc|cc}
Model & \multicolumn{2}{c|}{$\sigma(pp\rightarrow \tilde\nu_L\,\tilde \nu_L)$}& \multicolumn{2}{c}{$\sigma(pp\rightarrow \tilde\nu_L\,\tilde\ell_L)$}  \\
\hline
&14 TeV & 27 TeV & 14 TeV & 27 TeV\\
\hline \rule{0pt}{3ex}
\!\!(a) & 9.37 & 29.82 & 37.63  & 116.38   \\
 (b) & 2.80 & 10.25  & 11.76  & 41.56   \\
 (d) & 10.52 & 33.10 & 41.78  & 127.80   
\end{tabular}
\end{ruledtabular}
\end{table}

\begin{table}[t]
\caption{\label{tab5} The estimated integrated luminosities, in fb$^{-1}$, for discovery of benchmarks (a), (b) and (d) of Table~\ref{tab1} at 14 TeV and 27 TeV for the cases of sneutrino pair production and slepton associated production with a sneutrino.} 
\begin{ruledtabular}
\begin{tabular}{ccc|cc}
Model & \multicolumn{2}{c|}{SR-$2\ell1$j [$\tilde\nu_L\,\tilde \nu_L$]}& \multicolumn{2}{c}{SR-$2\ell1$j [$\tilde\nu_L\,\tilde\ell_L$]}  \\
\hline \rule{0pt}{3ex}
& $\mathcal{L}$ at 14 TeV & $\mathcal{L}$ at 27 TeV & $\mathcal{L}$ at 14 TeV & $\mathcal{L}$ at 27 TeV\\
\hline \rule{0pt}{3ex}
\!\!(a) & 367 & 87  & 257  & 39    \\
 (b) & 685 & 127  & 295  & 68  \\
  (d) & 317 & 65 & 232 & 32
\end{tabular}
\end{ruledtabular}
\end{table}

\subsection{Slepton associated production with a sneutrino}

Finally, we consider the associated production of a slepton with an sneutrino for the benchmarks (a), (b) and (d) and for the same reason discussed in the sneutrino pair production case. In addition, this production mode proceeds through the charged current and thus has a larger cross section as one can see from Table~\ref{tab4}. This leads to a lower integrated luminosity for discovery at both HL-LHC and HE-LHC as illustrated in Table~\ref{tab5}. Fig.~\ref{figdnn2} shows the distributions in the DNN response for benchmark (d) in the single jet signal region at 14 TeV and 27 TeV for the slepton associated production channel. 

Next, we discuss the systematic uncertainties associated with the signal and the background and their effect on the predicted integrated luminosities.

\section{5. Discovery significance for benchmarks \label{5}}

The integrated luminosity for a $5\sigma$ discovery is re-estimated after including the systematic uncertainties using the signal significance
\begin{align}
\sigma=\frac{S}{\sqrt{S+B+(\delta_S S)^2+(\delta_B B)^2}},
\label{significance}
\end{align}
where $\delta_S$ and $\delta_B$ are the systematic uncertainties in the signal and background estimates. The recommendations on systematic uncertainties (known as `YR18' uncertainties) published in the CERN's yellow reports~\cite{CidVidal:2018eel,Cepeda:2019klc} suggest an overall 20\% uncertainty in the background and 10\% in the SUSY signal. The bottom pads of each of the panels in Figs.~\ref{figdnn1} and~\ref{figdnn2} show the distribution in the signal significance of Eq.~(\ref{significance}) as a function of the cut on `DNN response'. We adopt a finer binning in the bottom pads as compared to the upper ones in order to properly show how the significance changes with the cut. We notice that a higher integrated luminosity is required after including the systematics but are still within the reach of HL-LHC and HE-LHC.

Next, we combine the different production channels discussed earlier to present the final integrated luminosities for discovery of the benchmarks of Table~\ref{tab1}. We show in Fig.~\ref{figlumi} the integrated luminosities for benchmarks (a)$-$(e) before and after including the `YR18' uncertainties and combining the different production channels at HL-LHC and HE-LHC. The signal regions shown are the ones which give us the best sensitivity for SUSY discovery. We also show the integrated luminosities for discovery of the benchmarks in Table~\ref{tab6} after including systematic uncertainties in the signal and background. Thus, benchmarks (a), (b) and (d) are discoverable with $\mathcal{L}\sim 180$ to $\sim 260$ fb$^{-1}$ at 14 TeV while the estimate drops to $\sim 40$ to $\sim 72$ fb$^{-1}$ at 27 TeV. In both cases, the most optimal signal region is SR-$2\ell1$j. For benchmark (c), $\mathcal{L}\sim 900$ fb$^{-1}$ is required at HL-LHC and $\sim 600$ fb$^{-1}$ at HE-LHC with SR-$2\ell2$j being the optimal signal region. Lastly, benchmark (e) can be discovered at HL-LHC with $\sim 1300$ fb$^{-1}$ while $\sim 780$ fb$^{-1}$ of integrated luminosity is needed at HE-LHC with SR-$2\ell1$j being the optimal signal region for discovery.    \\

One final remark regarding the LHC phenomenology in this analysis. Benchmarks (c) and (e) exhibit light charginos and second neutralinos with a considerable mass gap between those particles on one hand and the neutralino LSP on another. 
Thus here one should also consider electroweakino pair production, $\tilde\chi^0_2\tilde\chi^\pm_1$ and $\tilde\chi^+_1\tilde\chi^-_1$. However, in neither of those benchmarks  the charginos and the second neutralinos are 
 the NLSP and it is the stau which is the NLSP. Further, in benchmark (e), $\tilde\chi^0_2$ and $\tilde\chi^\pm_1$ are heavier than the sleptons and the staus. For this reason, the branching ratio to SFOS leptons is greatly reduced especially for benchmark (c) where the electroweakinos decay to staus which eventually decay to a tau and an LSP. Of course, a tau can decay leptonically but this branching ratio is suppressed in comparison to its hadronic decays. Despite the larger production cross sections, the overall $\sigma\times\text{BR}$ turns out to be smaller than the other production modes considered in this paper. Thus the electroweakinos do not constitute a strong discovery channel for 
the benchmarks discussed  here. The interested reader is directed to earlier works on SUSY discovery with electroweakino production~\cite{Aboubrahim:2017wjl}, including the clean three-lepton channel~\cite{Aboubrahim:2018bil}.

\begin{table}[t]
\caption{\label{tab6} The estimated integrated luminosities, in fb$^{-1}$, for discovery of benchmarks of Table~\ref{tab1} at 14 TeV and 27 TeV after combining all production channels and including systematics in the signal and background. } 
\begin{ruledtabular}
\begin{tabular}{ccc|cc}
Model & \multicolumn{2}{c|}{SR-$2\ell1$j}& \multicolumn{2}{c}{SR-$2\ell2$j}  \\
\hline \rule{0pt}{3ex}
& $\mathcal{L}$ at 14 TeV & $\mathcal{L}$ at 27 TeV & $\mathcal{L}$ at 14 TeV & $\mathcal{L}$ at 27 TeV\\
\hline \rule{0pt}{3ex}
\!\!(a) & 180 & 40 & 1863 & 950   \\
 (b) & 260 & 72 & 1720 & 1550   \\
 (c) & 3155 & 1060 & 935 & 600   \\
 (d) & 200 & 50 & 1860 & 715   \\
 (e) & 1287 & 786 & 1437 & 1175  \\
\end{tabular}
\end{ruledtabular}
\end{table}

\section{6. Constraints on CP phases from \amu \label{6}}

It is known that SUSY CP violating phases arising from the soft parameters can have significant
effect on $a_{\mu}^{\rm SUSY}$~\cite{Ibrahim:1999aj,Ibrahim:2001ym}.
 Here we discuss the phase dependence of the chargino contribution
 which is the dominant one, although the analysis is done including both the chargino and the neutralino exchange contributions. For the chargino exchange contribution the phases enter via the chargino mass matrix 
\begin{equation}
M_{C}=
\begin{pmatrix}
|m_2|e^{i\xi_2} & \sqrt{2}  m_W  \sin\beta \\
\sqrt 2 m_W \cos\beta & |\mu| e^{i\theta_{\mu}}\\
\end{pmatrix},
\end{equation}
 where $\theta_{\mu}$ is the phase of the Higgs mixing parameter
$\mu$,  and $\xi_2$ is  the phase of the SU(2) gaugino mass $m_2$.
The chargino contribution is given by~\cite{Ibrahim:1999aj}
\begin{equation}
\begin{aligned}
a^{\chi^{+}}_{\mu}&=
\frac{m_{\mu}\alpha_{EM}}{4\pi\sin^2\theta_W}
\sum_{i=1}^{2}\frac{1}{m_{\chi_i^+}}\text{Re}(\kappa_{\mu} U^*_{i2}V^*_{i1})
F_3\left(\frac{m^2_{\tilde{\nu}}}{m^2_{\tilde\chi_i^+}}\right) \\
&\hspace{-1cm}+\frac{m^2_{\mu}\alpha_{EM}}{24\pi\sin^2\theta_W}
\sum_{i=1}^{2}\frac{1}{m^2_{\chi_i^+}}
(|\kappa_{\mu} U^{*}_{i2}|^2+|V_{i1}|^2)
F_4\left(\frac{m^2_{\tilde{\nu}}}{m^2_{\tilde\chi_i^+}}\right),
\label{achargino}
\end{aligned}
\end{equation}
where the form factors are given by
\begin{equation}
\begin{aligned}
F_3(x)&=\frac{1}{(x-1)^3}(3x^2-4x+1-2x^2 \ln x), \\ 
 F_4(x)&=\frac{1}{(x-1)^4}(2x^3+3x^2-6x+1-6x^2 \ln x)\,.
\end{aligned}
\end{equation}
In Eq.~(\ref{achargino}), $U$ and $V$ are defined so that $U^* M_C V^{-1}=\text{diag}(m_{\tilde\chi^+_1},m_{\tilde\chi^+_2})$, where $U$ and $V$ are  unitary 
matrices, and where $\kappa_{\mu}={m_{\mu}}/{\sqrt{2} m_W \cos\beta}$. 

\begin{figure}[!htp]
\centering
\includegraphics[width=0.49\textwidth]{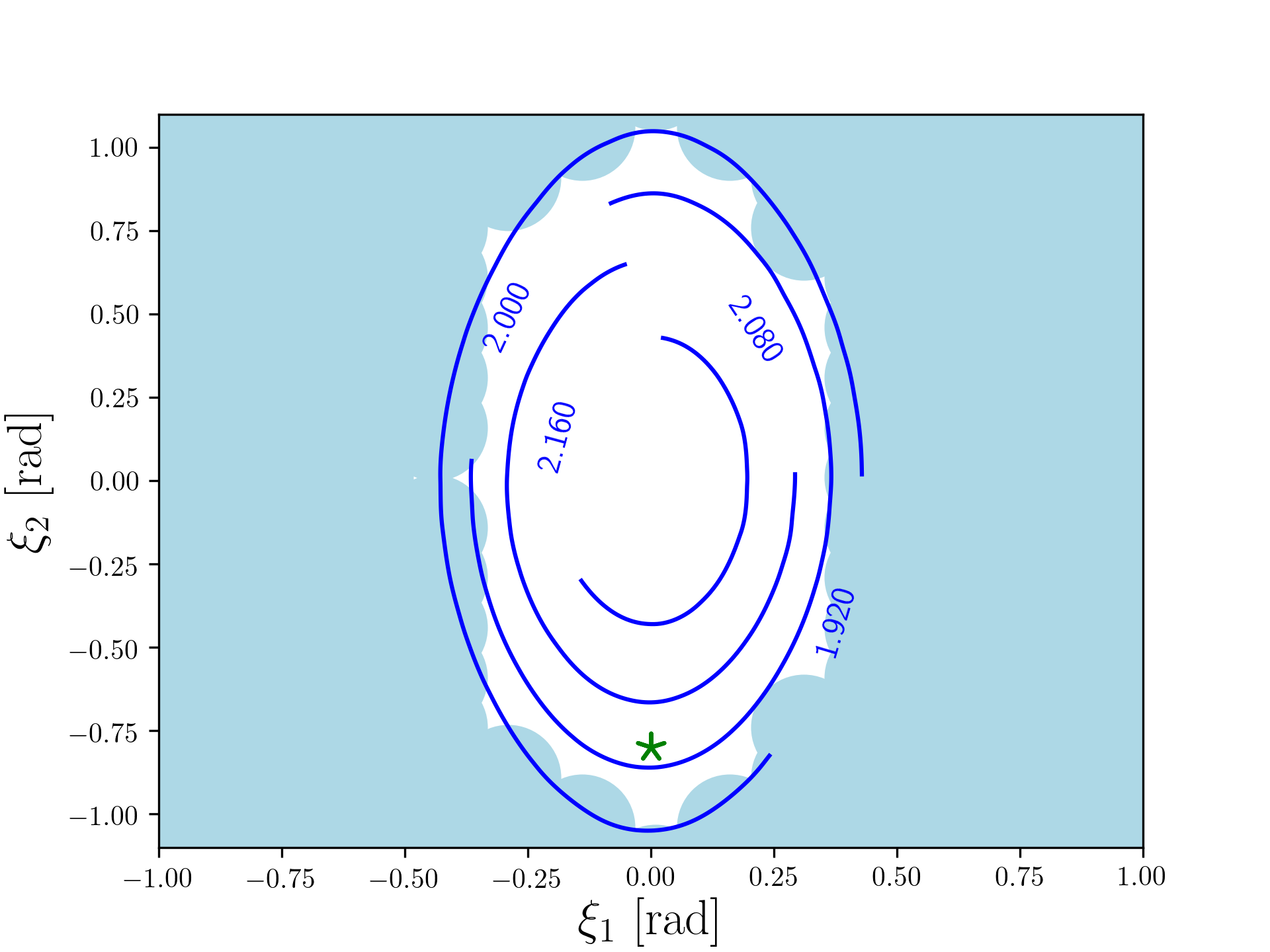}
\includegraphics[width=0.49\textwidth]{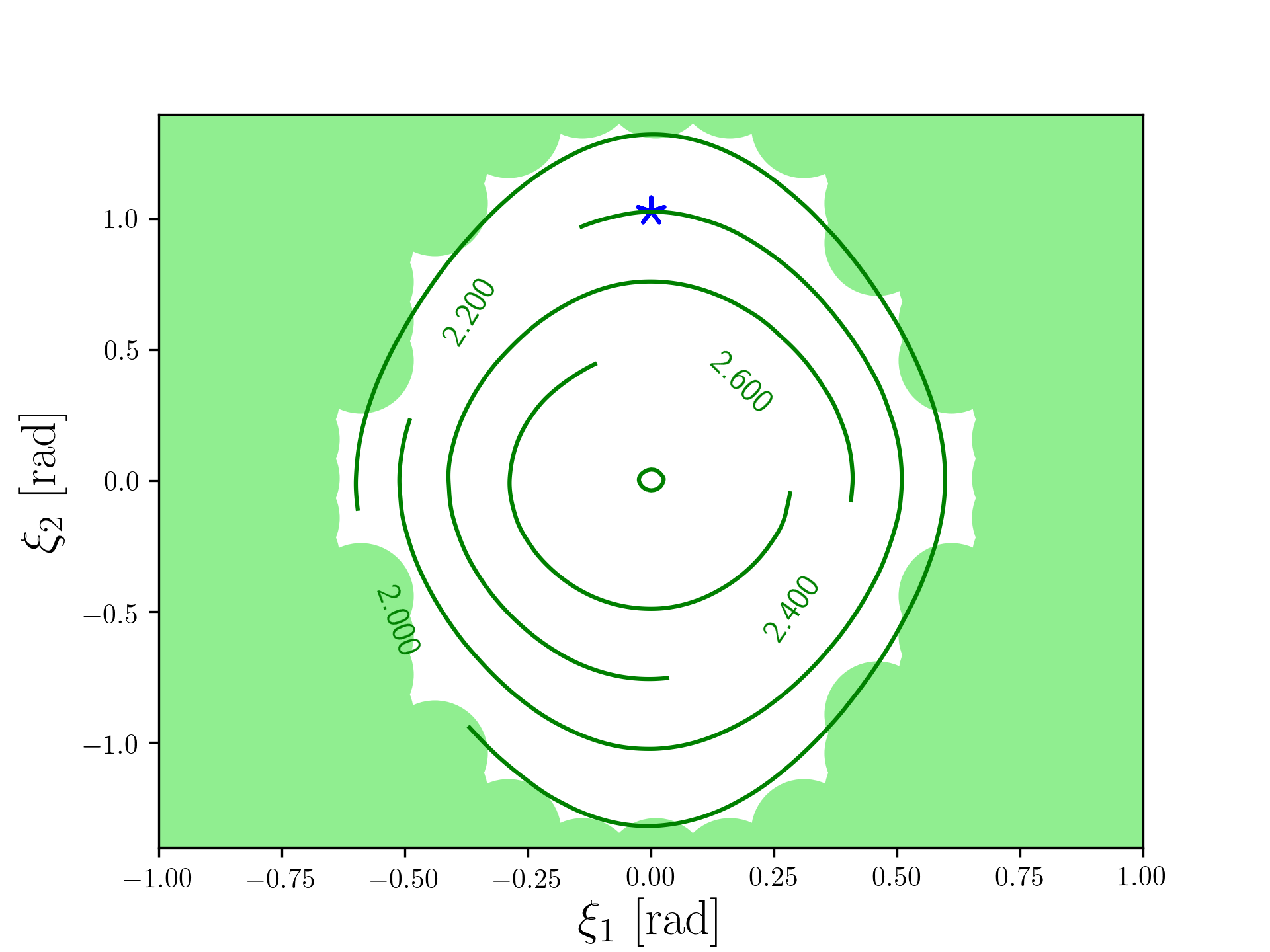}\\
\includegraphics[width=0.49\textwidth]{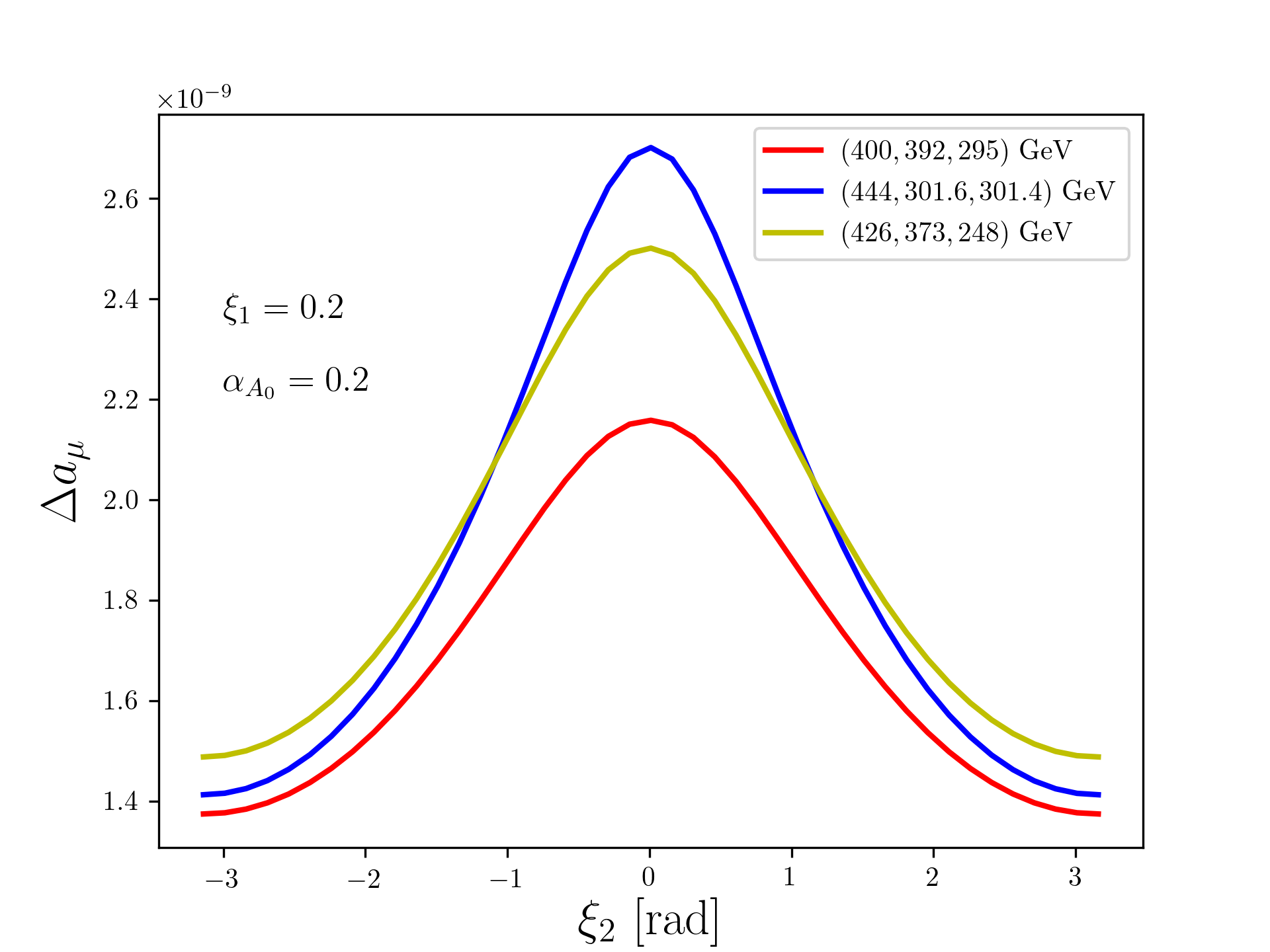}
\caption{
Top and middle panels: Exclusion plots in the CP phases $\xi_1$ and $\xi_2$ arising from the \amu~constraint. The contours correspond to $\Delta a_{\mu}\times 10^{9}$ consistent with the combined experimental result on $g-2$. Bottom panel: 
Variation of $a^{\rm SUSY}_\mu$ with the phase $\xi_2$. In all plots, $\alpha_{A_0}=0.2$ rad. The masses of the light particles in the loops are in this order $(\tilde\mu,\tilde\chi^{\pm}_1,\tilde\chi^0_1)$: $(400,392,1,391.7)$ GeV for the top panel and $(445,301.6,301.4)$ GeV for the middle panel and shown in the legend for the bottom panel. The star in the top and middle panels indicate illustrative  tiny regions of the cancellation mechanism~\cite{Ibrahim:1998je}
where one can simultaneously obtain a muon $g-2$ and the electron EDM consistent with experiment, $|d_e|<1.1\times 10^{-29}e$cm.} 
\label{figcp}
\end{figure}

The neutralino contribution is given by
\begin{align}
a^{\chi^{0}}_{\mu}&=\frac{m_{\mu}\alpha_{\rm EM}}{4\pi\sin^2\theta_W}\sum_{j=1}^4\sum_{k=1}^2\frac{1}{m_{\tilde\chi^0_j}}\text{Re}(\eta^k_{\mu j})F_1\left(\frac{m^2_{\tilde\mu_k}}{m^2_{\tilde\chi^0_j}}\right) \nonumber \\
&+\frac{m^2_{\mu}\alpha_{\rm EM}}{24\pi\sin^2\theta_W}\sum_{j=1}^4\sum_{k=1}^2\frac{1}{m^2_{\tilde\chi^0_j}}X^k_{\mu j}F_2\left(\frac{m^2_{\tilde\mu_k}}{m^2_{\tilde\chi^0_j}}\right),
\label{aneutralino}
\end{align}
where
\begin{align}
\eta^k_{\mu j}&=-\frac{1}{\sqrt{2}}\Big(\tan\theta_W X_{1j}D^*_{1k}+X_{2j}D^*_{1k}-\sqrt{2}\kappa_\mu X_{3j}D^*_{2k}\Big) \nonumber \\
&\hspace{0.8cm}\times \Big(\sqrt{2}\tan\theta_W X_{1j}D_{2k}+\kappa_\mu X_{3j}D_{1k}\Big),
\end{align} 
and
\begin{align}
X^k_{\mu j}&=|D_{1k}|^2\text{Re}(X_{1j}X^*_{2j})\tan\theta_W+\frac{m^2_{\mu}}{2m^2_W\cos^2\beta}|X_{3j}|^2 \nonumber \\
&+\frac{1}{2}\tan^2\theta_W |X_{1j}|^2(|D_{1k}|^2+4|D_{2k}|^2) \nonumber \\
&-\frac{m_{\mu}}{m_W\cos\beta}\text{Re}(X_{3j}X^*_{2j}D_{1k}D^*_{2k})+\frac{1}{2}|X_{2j}|^2|D_{1k}|^2 \nonumber \\
&+\frac{m_{\mu}\tan\theta_W}{m_W\cos\beta}\text{Re}(X_{3j}X^*_{1j}D_{1k}D^*_{2k}).
\label{xmu}
\end{align}
In Eq.~(\ref{xmu}), $X$ is a unitary matrix that diagonalizes the symmetric neutralino mass matrix, so that $X^T M_{\tilde\chi^0}X=\text{diag}(m_{\tilde\chi^0_1},m_{\tilde\chi^0_2},m_{\tilde\chi^0_3},m_{\tilde\chi^0_4})$, and $D$ diagonalizes the hermitian smuon mass square matrix, $D^{\dagger}M^2_{\tilde\mu}D=\text{diag}(m^2_{\tilde\mu_1},m^2_{\tilde\mu_2})$. The form factors in Eq.~(\ref{aneutralino}) are given by
\begin{equation}
\begin{aligned}
F_1(x)&=\frac{1}{(x-1)^3}(1-x^2+2x\ln x), \\ 
 F_2(x)&=\frac{1}{(x-1)^4}(-x^3+6x^2-3x-2-6x\ln x)\,.
\end{aligned}
\end{equation}
The phases enter via $U,V,X$ and through the chargino and neutralino masses.
We note that the phase 
dependence of the chargino contribution to $a_{\mu}$ arises entirely from the combination $\theta_{\mu}+\xi_2$. 
The neutralino contribution, however,  has additional phase dependence from 
 $\xi_1$, the phase of $m_1$, and from 
$\alpha_{A_{0}}$,  the phase of  $A_0$.
 In Fig.~\ref{figcp} we show the sensitivity of \amu~to CP phases $\xi_1$ and $\xi_2$.

The top panels of Fig.~\ref{figcp} show the excluded regions (shaded) in the $\xi_1$-$\xi_2$ plane due to the \amu~constraint for two points chosen from the large set of points obtained from the scan. The contours shown in the allowed regions correspond to $\Delta a_{\mu}$ consistent with the combined Brookhaven and Fermilab results. 
The lower panel shows the sensitivity of $\Delta a_{\mu}$ to the CP phase $\xi_2$ with a range of values consistent with the recent $(g-2)_{\mu}$ experiment. 
We note here that in addition to the constraint on the CP phases by the  $(g-2)_{\mu}$ experiment, 
the phases are also subject to the EDM constraints. Thus while the phases satisfying the EDM constraints must lie
in the white regions of the upper two panels of Fig.~\ref{figcp}, the EDM constraints on them are much stronger,
the strongest being the electron EDM which has the upper limit of $|d_e| <1.1\times 10^{-29} e$cm.
 For models with low slepton  mass spectrum, which is the case here, a satisfaction of the EDM constraint
 can come about via the cancellation mechanism~\cite{Ibrahim:1998je} in tiny regions of the parameter space. 
 An illustration of this phenomenon is given in the upper two panels of Fig.~\ref{figcp} where we display two tiny regions of the
 parameter space where the electron EDM constraint is satisfied.

\section{7. Conclusion \label{7}}
  In this work we have shown that the combined Fermilab and Brookhaven data on 
  $a^{\rm exp}_\mu-a^{\rm SM}_\mu$ has  
  important   implications  for the discovery of supersymmetry at HL-LHC and HE-LHC.
   Specifically, exploration of the 
  SUGRA parameter space using machine learning shows that the combined
   Fermilab and Brookhaven  \amu~constraint indicates
  that the favored region of the  parameter space is that of $\tilde g$SUGRA where gluino-driven radiative breaking of the
 electroweak symmetry occurs. In this region the renormalization group
  analysis leads to a split light and heavy mass spectrum 
 where the electroweak gauginos and the sleptons are light lying in the few hundred GeV range, while the
  remaining mass spectrum is heavy. The light spectrum which includes the neutralino, the chargino,  
  the smuon, and the smuon-neutrino can produce a correction to the muon anomaly consistent with 
  \amu. Further, the light stau and the chargino are seen to be the lightest charged
  particles   while the sleptons are light enough to be prime
  candidates for discovery at HL-LHC and HE-LHC. We perform a signal region analysis and compute the integrated luminosity needed for SUSY discovery. 
  It is shown that supergravity  models which produce a correction to $\Delta a^{\rm SM}_{\mu}$ of size indicated \amu~are discoverable at HL-LHC within the optimal integrated luminosity and with a smaller integrated luminosity 
  at HE-LHC. It is also shown that \amu~puts constraints on the CP phases that enter the muon anomaly and eliminates 
   significant regions of their parameter space. These constraints are independent of 
   the EDM constraints which must be imposed in the regions of CP phases allowed by the muon anomalous magnetic moment. Finally we note here some previous and recent works related to the $g-2$ anomaly~\cite{works}. \\~\\
  
 \begin{acknowledgments}
 The research of AA and MK was supported by the BMBF under contract 05H18PMCC1. The research of PN  was supported in part by the NSF Grant PHY-1913328.
\end{acknowledgments}

\end{document}